\documentclass[fleqn,usenatbib]{mnras}

\usepackage{newtxtext,newtxmath}
\usepackage{mathtools}
\usepackage[T1]{fontenc}

\DeclareRobustCommand{\VAN}[3]{#2}
\let\VANthebibliography\thebibliography
\def\thebibliography{\DeclareRobustCommand{\VAN}[3]{##3}\VANthebibliography}

\newcommand{\kep}{{\it Kepler}}

\newcommand{\sph}{$S_\text{\!ph}$}

\newcommand{\tacf}{$\tau_\text{ACF}$}

\newcommand{\avtacf}{$\tilde{\tau}_\text{ACF}$}
\newcommand{\tinput}{$\tau_\text{input}$}

\usepackage{colortbl}
\definecolor{purple}{RGB}{102,0,204}
\definecolor{green}{RGB}{51,102,0}

\usepackage{graphicx}	% Including figure files
\usepackage{amsmath}	% Advanced maths commands
\title[Active-region lifetimes and the ACF]{On the relation between active-region lifetimes and the autocorrelation function of light curves}

\author[A. R. G. Santos et al.]{
A. R. G. Santos,$^{1,2}$\thanks{\!\!\!E-mail: angela.goncalves-dos-santos@warwick.ac.uk; asantos@astro.up.pt} %0000-0001-7195-6542
S. Mathur,$^{3,4}$ %0000-0002-0129-0316
R. A. Garc\'{i}a,$^{5}$ %0000-0002-8854-3776
M. S. Cunha,$^{6}$ %0000-0001-8237-7343
and P. P. Avelino, $^{6,7}$ %0000-0002-1440-6963
\\
$^{1}$Department of Physics, University of Warwick, Coventry, CV4 7AL, UK\\
$^{2}$Space Science Institute, 4765 Walnut Street, Suite B, Boulder CO 80301, USA\\
$^{3}$Instituto de Astrof\'isica de Canarias (IAC), E-38205 La Laguna, Tenerife, Spain\\
$^{4}$Universidad de La Laguna (ULL), Departamento de Astrof\'isica, E-38206 La Laguna, Tenerife, Spain\\
$^{5}$AIM, CEA, CNRS, Universit\'e Paris-Saclay, Universit\'e de Paris, Sorbonne Paris Cit\'e, F-91191 Gif-sur-Yvette, France\\
$^{6}$Instituto de Astrof\'{i}sica e Ci\^{e}ncias do Espa\c{c}o, Universidade do Porto, CAUP, Rua das Estrelas, PT-4150-762 Porto, Portugal\\
$^{7}$Departamento de F\'{i}sica e Astronomia, Faculdade de Ci\^{e}ncias, Universidade do Porto, Rua do Campo Alegre 687, PT-4169-007 Porto, Portugal
}

\date{Accepted XXX. Received YYY; in original form ZZZ}

\pubyear{2021}

\begin{document}
\label{firstpage}
\pagerange{\pageref{firstpage}--\pageref{lastpage}}
\maketitle

\begin{abstract}
Rotational modulation of stellar light curves due to dark spots encloses information on spot properties and, thus, on magnetic activity. In particular, the decay of the autocorrelation function (ACF) of light curves is presumed to be linked to spot/active-region lifetimes, given that some coherence of the signal is expected throughout their lifetime. In the literature, an exponential decay has been adopted to describe the ACF. Here, we investigate the relation between the ACF and the active-region lifetimes. For this purpose, we produce artificial light curves of rotating spotted stars with different observation, stellar, and spot properties. We find that a linear decay and respective timescale better represent the ACF than the exponential decay. We therefore adopt a linear decay. The spot/active-region timescale inferred from the ACF is strongly restricted by the observation length of the light curves. For 1-year light curves our results are consistent with no correlation between the inferred and the input timescales. The ACF decay is also significantly affected by differential rotation and spot evolution: strong differential rotation and fast spot evolution contribute to a more severe underestimation of the active-region lifetimes. Nevertheless, in both circumstances the observed timescale is still correlated with the input lifetimes. Therefore, our analysis suggests that the ACF decay can be used to obtain a lower limit of the active-region lifetimes for relatively long-term observations. However, strategies to avoid or flag targets with fast active-region evolution or displaying stable beating patterns associated with differential rotation should be employed.
\end{abstract}
% of the light curve of solar-type stars %and analyse
%For simple signals, the ACF decay is linear rather than exponential. For more complex signals, the linear decay is till a better representation of the ACF decay.


%%%%%%%%%%%%%%%%%%%%%%%%%%%%%%%%%%%%%%%%%%%%%%%%%%

\section{Introduction}

Dark magnetic spots emerge at the surface of active low-mass stars with convective outer layers, i.e. stars from spectral type F to M (hereafter solar-type stars). As the star rotates, the dark spots at its surface modulate the stellar brightness. Such modulation encloses important information on stellar magnetic activity and rotation properties.

%The advent of space-based planet-hunting missions provided the unique opportunity for studying these properties. In particular,
During its nominal mission, the \kep\ satellite \citep{Borucki2010} collected high-precision long-term and continuous light curves for more than 100,000 solar-type stars. For those light curves exhibiting spot modulation, it has been possible to constrain activity-related properties, such as average surface rotation \citep[e.g.][]{McQuillan2014,Mathur2014,Garcia2014,Santos2019a,Santos2021,Godoy-Rivera2021}, differential rotation \citep[e.g.][]{Reinhold2013a,Karoff2018}, and average photometric magnetic activity \citep[e.g.][]{Mathur2014,Garcia2014,Santos2019a,Santos2021}. Another property that, in principle, can be constrained from the spot modulation is the lifetimes of active regions.

The decay timescale of the autocorrelation function of light curves is expected to be related to the active-region lifetimes \citep[e.g.][]{Lanza2014}. \citet{Giles2017} measured the ACF $e$-folding time for about 2,200 \kep\ solar-type stars, using this timescale as an estimate of the active-region lifetime. The authors concluded that stars with larger brightness variations have longer-lived active regions in comparison to stars with smaller brightness variations, and that active-region lifetime decreases with effective temperature.

In this work, we explore the relation between the decay of the autocorrelation function and the active-region lifetimes. To that end we use artificial light curves obtained with different properties, such as observation length, surface differential rotation, and spot evolution rates. The properties of the artificial data are detailed in Section~\ref{sec:data}. Section~\ref{sec:method} describes the model for the autocorrelation function adopted by \citet{Giles2017} and the new parameterization we propose. In particular, we argue that the ACF decay is linear and that one cannot recover the input lifetimes through the $e$-folding time. In Section~\ref{sec:restacf}, we present the results for the different sets of artificial data and, in Section~\ref{sec:conclusion}, we draw our conclusions.

\section{Artificial light curves}\label{sec:data}

In this work, we use artificial data to test the autocorrelation timescale, \tacf, in order to determine how it relates with the active-region lifetime.

Starspots are generated using the tool developed by \citet{Santos2015}, who successfully applied it to the sunspot cycle. The tool requires a set of input parameters (see below), which can be adapted to the target spot and stellar properties. The time-dependent area and location of the generated spots are then adopted by the light-curve generation tool.

The artificial light curves are obtained by the tool developed in \citet{Santos2017a}. The relative decrease in flux due to spots depends on the relative spot area, spot location, stellar inclination, limb-darkening, and spot-to-photosphere intensity ratio. In this work, for simplicity the limb-darkening law and spot-to-photosphere contrast $C_\text{S}$ are fixed: $C_\text{S}=0.67$ -- fixed at the solar value; quadratic limb-darkening law with parameters $a_1=0.5287$ and $a_2=0.2176$ -- adequate for solar-type stars \citep[e.g.][for more details see \citet{Santos2017a}]{Sofia1982,Claret2000}.

In what follows, we describe the remainder of the input parameters. Particularly, we list the values adopted for the reference artificial light curves (see also Table~\ref{tab:parameters}). In Section~\ref{sec:restacf}, we explore the impact of the different parameters on the ACF decay. For a detailed implementation of the parameters see \citet{Santos2015}.

{\it Observation length and cadence.} The cadence is set to two hours, which is the cadence to which we re-bin the \kep\ long-cadence data (30 min) for the rotation analysis \citep[e.g.][]{Mathur2010}. The reference light curves have an observation length, $t_\text{obs}$, of 4 years to be conformant with \kep\ data. In Section~\ref{sec:length}, we vary the length of the light curves to shorter observation lengths in order to study its effect on the ACF decay.

{\it Number of spots.} At each time step, spots are randomly generated according to a Poisson distribution with a mean value, which is the input parameter. For the reference light curves we keep the number of spots small (typically less than 10) to reduce computing time. Note that for the reference light curves the number of spots is variable as they are randomly generated. In Section~\ref{sec:snumber}, we test the effect of the number of spots on the ACF decay, by fixing the number of allowed spots and also by considering a higher probability for the spot formation.

{\it Stellar inclination.} The stellar inclination angle $i$ is the angle between the rotation axis and the line of sight. For the reference light curves, $i$ is fixed at $70^\circ$. In Section~\ref{sec:inc} we investigate the impact of the inclination angle.

{\it Spot location.} Except for the active longitude exercise (Section~\ref{sec:longitudes}), initial spot longitudes are randomly taken according to a uniform distribution. For the reference light curves, spot latitudes $L$ are also random according to a Gaussian distribution with mean $\langle L\rangle$ and standard deviation $\sigma_\text{L}$ fixed at $15^\circ$ and $5^\circ$, respectively. In Section~\ref{sec:rot}, the input parameters for the spot latitudes, i.e. the spot formation zone, are varied to assess their impact on the ACF decay.

{\it Spot size and lifetime.} For the reference light curves, the spot maximum area is fixed at a given value, i.e. all spots in a given simulation have the same maximum area (hereafter maximal area to avoid confusion in Section~\ref{sec:vA}). In the Sun, the lifetimes of sunspots and sunspot groups are known to be proportional to their maximal area, as described by the Gnevyshev-Waldmeier rule \citep[][]{Gnevyshev1938,Waldmeier1971}: $A_\text{maximal}=D_\text{GW}\tau$, where $A_\text{maximal}$ is the spot maximal area in millionth of the solar hemisphere ($\mu\text{Hem}$), $D_\text{GW}$ is the constant of proportionality, and $\tau$ is the lifetime. In this work, we adopt the Gnevyshev-Waldmeier rule and the constant $D_\text{GW}=10\,\mu\text{Hem\,day}^{-1}$ for our artificial light curves of spotted stars. We note that different authors found slightly different values around $10\,\mu\text{Hem\,day}^{-1}$ for $D_\text{GW}$ \citep[e.g.][]{Petrovay1997,Henwood2010}. These differences arise from the difficulty of measuring lifetimes of individual sunspots or sunspot groups due to nightfall, surface rotation, and limb darkening. In Section~\ref{sec:vA}, we vary the spot areas and respective lifetimes within the same simulation, where random areas are drawn from a log-normal distribution. In the Sun, sunspot growth and decay rates -- change in spot area -- are described by a power law of the form $\Gamma=e^{\gamma_1} A^{\gamma_2}$ \citep[e.g.][]{Petrovay1997,Javaraiah2012}. Hereafter for simplification we refer to $\gamma_2$ as $\gamma$, and we consider $e^{\gamma_1}=1$. For the reference data set, the exponent $\gamma$ is fixed at 0.2 and, in Section~\ref{sec:evolution}, by varying $\gamma$, we explore the impact of spot evolution on the ACF decay.

{\it Surface rotation profile.} For the reference light curves, we assume a rotation profile similar to the solar surface rotation: $\Omega=\Omega_\text{eq}(1-\alpha\sin^2L)$, where $\Omega=2\pi/P$ is the angular velocity, $P$ is the period, ``eq" denotes equator, and $\alpha$ describes the shear. For the reference simulations $P_\text{eq}=25$ days and $\alpha=0.2$. In Section~\ref{sec:rot}, we explore different rotation profiles.

%\pagebreak

\section{Autocorrelation timescale}\label{sec:method}

The autocorrelation function (ACF) of a light curve with rotational modulation can be described as an underdamped harmonic oscillator with an interpulse term \citep{Giles2017}: 
\begin{equation}
    y(t)=e^{-\dfrac{t}{\tau_e}}\left[a\cos\left(\dfrac{2\pi t}{P_\text{ACF}}\right)+b\cos\left(\dfrac{4\pi t}{P_\text{ACF}}\right)+y_0\right],
    \label{eq:fit0}
\end{equation}
where $t$ is the temporal lag, $\tau_e$ is the $e$-folding time of the ACF, $P_\text{ACF}$ is the rotation period inferred from the ACF, and $a$, $b$, and $y_0$ are constants. The $e$-folding time has been used as an estimate of the lifetime of the active-regions responsible for the modulation. The interpulse term is necessary to describe the ACF when the rotational signal is caused by spots or active regions that are apart in longitude by approximately $180^\circ$.

However, does $\tau_e$ stand for the spot/active-region lifetimes? 

In order to answer this question, we produce a simple set of 500 one-spot light curves. While the spot signature is very simple, part of the spot parameters are random, to be more compatible with the light curves of spotted stars below. In each realization only one spot is allowed to be formed over the 4 years of observation and its area is kept constant, i.e. the spot does not evolve and the amplitude of the spot modulation is the same throughout its lifetime. The spot emergence is random in time as well as its latitude and initial longitude, while ensuring that the spot emerges and dissipates within the observation time. The spot latitude, in this exercise, is drawn from a uniform distribution between $20^\circ$ and $80^\circ$ ($i=70^\circ$). The area of the spot is also random (only one spot per light curve) drawn from a uniform distribution between 1000 and 4000 $\mu\text{Hem}$, i.e. \tinput\ varies between 100 and 400 days. Since spots can emerge/dissipate at the far-side of the star, the spot location both in longitude and latitude slightly affects \tacf. In particular, the higher the latitude the longer the spot is visible which reduces the chances of the spot's signature not being detected throughout its full lifetime.

In addition, we obtain another one-spot light curve, where the various parameters are fixed. The spot is formed at latitude $L=25^\circ$ ($i=70^\circ$) and longitude $-90^\circ$, i.e. the spot emerges at the limb and starts moving across the near-side of the star. After ten rotations, the spot dissipated at the limb on the opposite side to where it first emerged. This way, the spot lifetime corresponds to the length of the observed spot modulation, i.e. the spot does not emerge or dissipate at the far-side of the star. The spot does not evolve (constant area) and its lifetime is then $\tau_\text{input}=10\times P+t_\text{vis}$, where $P$ stands for rotation period and $t_\text{vis}$ corresponds to the time that the spot is visible during a single rotation. At $L=25^\circ$, the rotation period is about 25.926 days ($P_\text{eq}=25$ days; $\alpha=0.2$) and $t_\text{vis}\sim15$ days, as follows $\tau_\text{input}\sim274.26$ days. 

\begin{figure}
    \centering
    \includegraphics[width=\hsize]{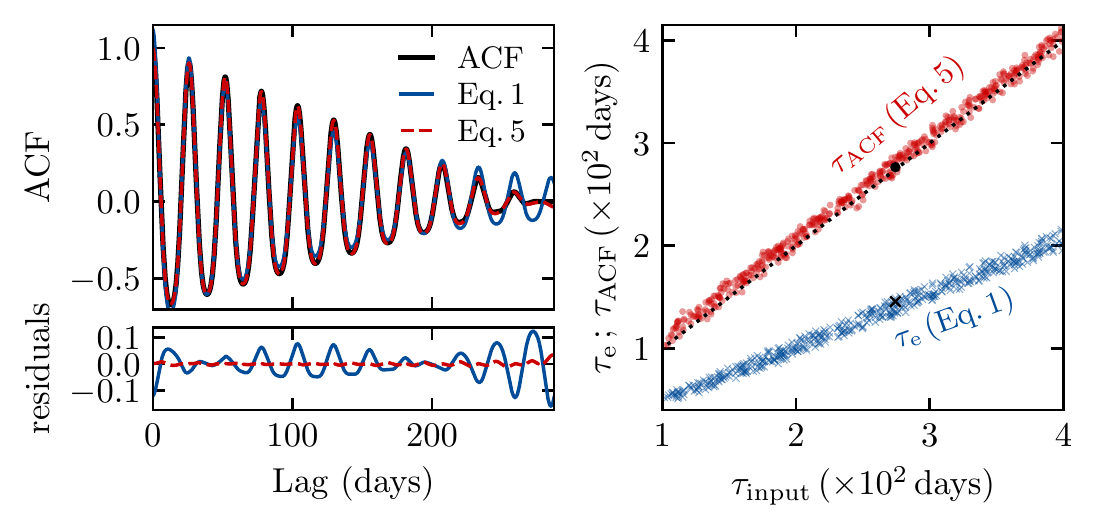}
    \caption{{\it Top left:} ACF (solid black) of the one-spot light curve with fixed parameters, where it is ensured that the spot lifetime corresponds to the length of the detected spot modulation (see text). The dashed red line shows the best fit with Eq.~\ref{eq:fit}, while the solid blue line shows the best fit with Eq.~\ref{eq:fit0}. {\it Bottom left:} Residuals between the ACF and its best fits with Eq.~\ref{eq:fit} (dashed red) and Eq.~\ref{eq:fit0} (solid blue). {\it Right:} \tacf (red circles) and $\tau_e$ (blue crosses) as a function of \tinput\ for one-spot light curves with random spot parameters. The black symbols highlight the respective results for the one-spot light curve with fixed parameters (that in the left). The black dotted line marks the 1-1 line.}
    \label{fig:1-spot}
\end{figure}

For each light curve, we then compute and fit the ACF with Eq.~\ref{eq:fit0} (see details in Sect.~\ref{sec:fit}). Even for simple spot signatures with constant maximum flux variation, we find that $\tau_e$ is only about half of the input lifetime (blue crosses in the right-hand side panel of Figure~\ref{fig:1-spot}).

How does the amplitude of the ACF decay? While the $e$-folding time of the ACF has been taken as a measure of the lifetimes of active regions as aforementioned, here we compute analytically how the amplitude of the ACF decays.

The ACF at lag $t_j$ can be written as 
\begin{equation}
    \text{ACF} (t_j)=\dfrac{1}{\int_{0}^{t_\text{obs}} f^2(t)\text{d}t}\int_{0}^{t_\text{obs}} f(t+t_j) f(t) \text{d}t,\label{eq:ACF}
\end{equation}
where $f$ is the flux variation, $\int_{0}^{t_\text{obs}} f^2(t)\text{d}t$ is the normalizing constant, and $t_\text{obs}$ is the observation length. Assuming a simple constant flux variation of the form
\begin{equation}
f(t)= \begin{cases}
  1, & \text{if}\,\, t\leq\tau,\\
  0, & \text{if}\,\, t>\tau,
\end{cases}
\end{equation}
the ACF for this simple signal becomes
\begin{equation}
    \text{ACF} (t_j)=1-\dfrac{t_j}{\tau},\label{eq:decay}
\end{equation}
where the second integral in Eq.~\ref{eq:ACF} (unnormalized ACF) is $\tau-t_j$, $\tau$ being the lifetime. The amplitude of the ACF varies linearly, and Eq.~\ref{eq:decay} defines the ACF decay in more general cases, e.g. periodic spot modulation with constant maximum flux amplitude. However, the rotational modulation is more complex with multiple evolving spots or active regions contributing to the signal. Nevertheless, as we will show in Section~\ref{sec:restacf}, the linear decay timescale is significantly closer to the input values than the $e$-folding time even for more complex signals.
One may notice that Eq.~\ref{eq:decay} corresponds to the two first terms of the series expansion of $e^{-t/\tau}$ (Eq.~\ref{eq:fit0}) and, hence, $\tau_e$ is still a constraint on the lifetime, while notably underestimating it.

We thus propose a new parameterization with a linear decay
\begin{equation}
    y(t)={\left(1-\dfrac{t}{\tau_\text{ACF}}\right)}\left[a\cos\left(\dfrac{2\pi t}{P_\text{ACF}}\right)+b\cos\left(\dfrac{4\pi t}{P_\text{ACF}}\right)+y_0\right],
    \label{eq:fit}
\end{equation}
where $\tau_\text{ACF}$ is the decay timescale of the ACF, which corresponds to the observed spot or active-region lifetime. 

The left panel of Figure~\ref{fig:1-spot} compares the best fits (see Sect.~\ref{sec:fit}) with Eq.~\ref{eq:fit0} (solid blue) and Eq.~\ref{eq:fit} (dashed red) to the ACF of the one-spot light curve above with fixed parameters. Eq.~\ref{eq:fit} is better suited to describe the ACF of a light curve than Eq.~\ref{eq:fit0}. The spot lifetime for that particular light curve is $\tau_\text{input}\sim274.26$ days. The recovered spot timescale \tacf\ is $\sim 276.49_{-0.22}^{+0.23}$ days, while $\tau_e\sim145.62_{-0.32}^{-0.29}$ days. The results for this light curve are highlighted in black in the right-hand panel of Figure~\ref{fig:1-spot}. The other symbols show the results for the remaining 500 one-spot light curves with random parameters. As described above the $e$-folding time is only about half of the input lifetimes, while \tacf\ closely follows the 1-1 line.

For the remainder of this work (Section~\ref{sec:restacf}), we adopt Eq.~\ref{eq:fit} to describe the ACF of the light curves.

\pagebreak

\subsection{Fitting method}\label{sec:fit}

For each artificial light curve, we obtain the ACF using the NumPy's function \texttt{correlate} and fit its full decay phase with Eq.~\ref{eq:fit0} (only Figures~\ref{fig:1-spot} and \ref{fig:stdsimul}) and Eq.~\ref{eq:fit}. We determine the decay phase of the ACF by finding when the amplitude of the consecutive ACF peaks, corresponding to multiples of $P_\text{ACF}$, ceases to decrease systematically (at least for two consecutive peaks). The length of the decay phase varies from light curve to light curve, being the longest for the longest input lifetimes.

At larger lags the ACF is evaluated using fewer data points than at smaller lags where most of the time series is accounted for (see Eq.~\ref{eq:ACF}). Therefore, we consider an uncertainty associated to each ACF value that is inversely proportional to the square root of the number of data points used, $N_\text{Flux}$, i.e. $\sigma_j=1/\sqrt{N_{\text{Flux},j}}$, where $j$ indicates a given lag. We note that this effect does not significantly change the fit.

The fit to the ACF is performed through the implementation of \texttt{emcee} \citep{Foreman-Mackey2013}, built upon the Affine Invariant Markov Chain Monte Carlo (MCMC) Ensemble sampler \citep{Goodman2010}. The logarithm of the likelihood function is given by
\begin{equation}
    \ln \mathcal{L}=-\dfrac{1}{2}\sum_{j=0}^{N_\text{ACF}}\left[\left(\dfrac{\text{ACF}_j-y_j}{\sigma_j}\right)^2+\ln\left(2\pi\sigma_j^2\right)\right],
    \label{eq:Likelihood}%least squares solution
\end{equation}
where $N_\text{ACF}$ is the number of ACF data points considered in the fit, i.e. the full decay phase described above. In the discrete case, the ACF (Eq.~\ref{eq:ACF}) can be written as
\begin{equation}
    \text{ACF}_j=\dfrac{1}{\sum_k f_k^2}\sum_k f_{j+k}\times f_k,\label{eq:ACF2}
\end{equation}
which, as mentioned above, we compute through NumPy's function \texttt{correlate}, and $k=0,\cdots,N_\text{ACF}$.

We adopt ignorance priors for all parameters in Eq.~\ref{eq:fit}, except for the rotation period. This type of analysis would be applied to light curves of stars known to exhibit rotational modulation, i.e. stars with known rotation periods. In the following paragraphs we describe the adopted prior functions.

{\it Observed spot timescale,} \tacf. We use a Jeffrey's prior for \tacf, which is appropriate for parameters that can range orders of magnitude. The minimum and maximum values are 0.1 and 2000 days (the longest input lifetime is 1000 days). The exception is for the exercise in Sect.~\ref{sec:vA}, where we consider the maximum to be 5000 days as spots lifetimes are randomly drawn. For Figures~\ref{fig:1-spot} and \ref{fig:stdsimul}, we use the same priors for $\tau_e$.

{\it Rotation period,} $P_\textrm{ACF}$. For this parameter in particular, we can take advantage of prior knowledge to define the prior function. We adopt a Gaussian distribution with mean value corresponding to $P_\text{eq}$ and a standard deviation being 20\% of $P_\text{eq}$. For example, on average the uncertainty on the rotation period from the wavelet analysis for \kep\ targets is 10\% \citep{Santos2019a,Santos2021}. A standard deviation of $20\%$ is also sufficient to account for the differential rotation adopted in this work.

{\it Constants $a$, $b$, $y_0$.} We adopt uniform distributions for these constants. The minimum and maximum values for $a$ are 0.01 and 1.5, for $b$ are 0.001 and 1, and for $y_0$ is -1 and 1. $b$ is typically smaller than $a$ as it is related to the amplitude of the interpulse term.

\texttt{emcee} uses an ensemble of interacting walkers to explore the parameter space. We adopt 25 walkers and a burn-in phase of 8000 steps after which each chain runs for 2000 steps. 
The final parameter estimates and the respective uncertainty are based on the median and the 68\% credible region of the marginalized posterior distribution.

\begin{figure}%%[h]
    \centering
    \includegraphics[width=\hsize]{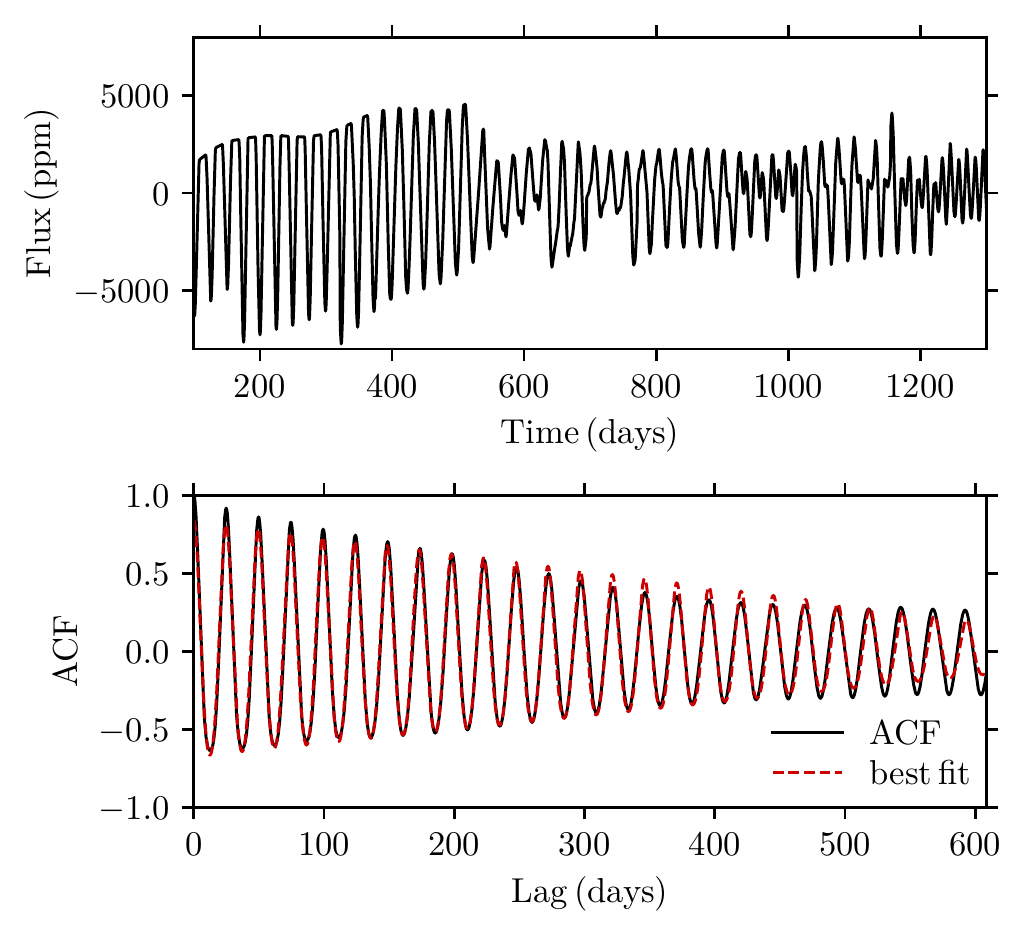}
    \caption{{\it Top:} Example of an artificial reference light curve, which is normalized in a similar way to the \kep\ data \citep[e.g.][]{Garcia2011}. {\it Bottom:} Autocorrelation function of the light curve (solid black) and its best fit (dashed red). The properties of the light curve are: $t_\text{obs}=4$ years; $i=70^\circ$; $\langle L\rangle=15^\circ$ and $\sigma_\text{L}=5^\circ$; input lifetime $\tau_\text{input}=600$ days; $\gamma=0.2$; $P_\text{eq}=25$ days and $\alpha=0.2$; 16 randomly generated spots.}
    \label{fig:ACFfit}
\end{figure}

Figure~\ref{fig:ACFfit} shows an example of the ACF and its best fit with Eq.~\ref{eq:fit}; the respective light curve is part of the reference data set described in Section~\ref{sec:data}. For easy access, Table~\ref{tab:parameters} summarizes the parameters that are often used in this work, as well as the values adopted for the reference data set.

\pagebreak

\section{Results}\label{sec:restacf}

In Section~\ref{sec:method}, for simple signals we demonstrate that the timescale of a linear decay of the ACF corresponds to the lifetime. However, the light curves of solar-type stars are unlikely to be consistent with one-spot light curves with unchanged amplitude in time.
The rotational modulation in the light curves of solar-type stars depends on a number of different observational, stellar, and active-region properties. In order to determine whether the lifetimes of active regions can be constrained from the ACF of light curves, we carry out a number of control tests with artificial data.

\begin{figure}%[h]
    \centering
    \includegraphics[width=\hsize]{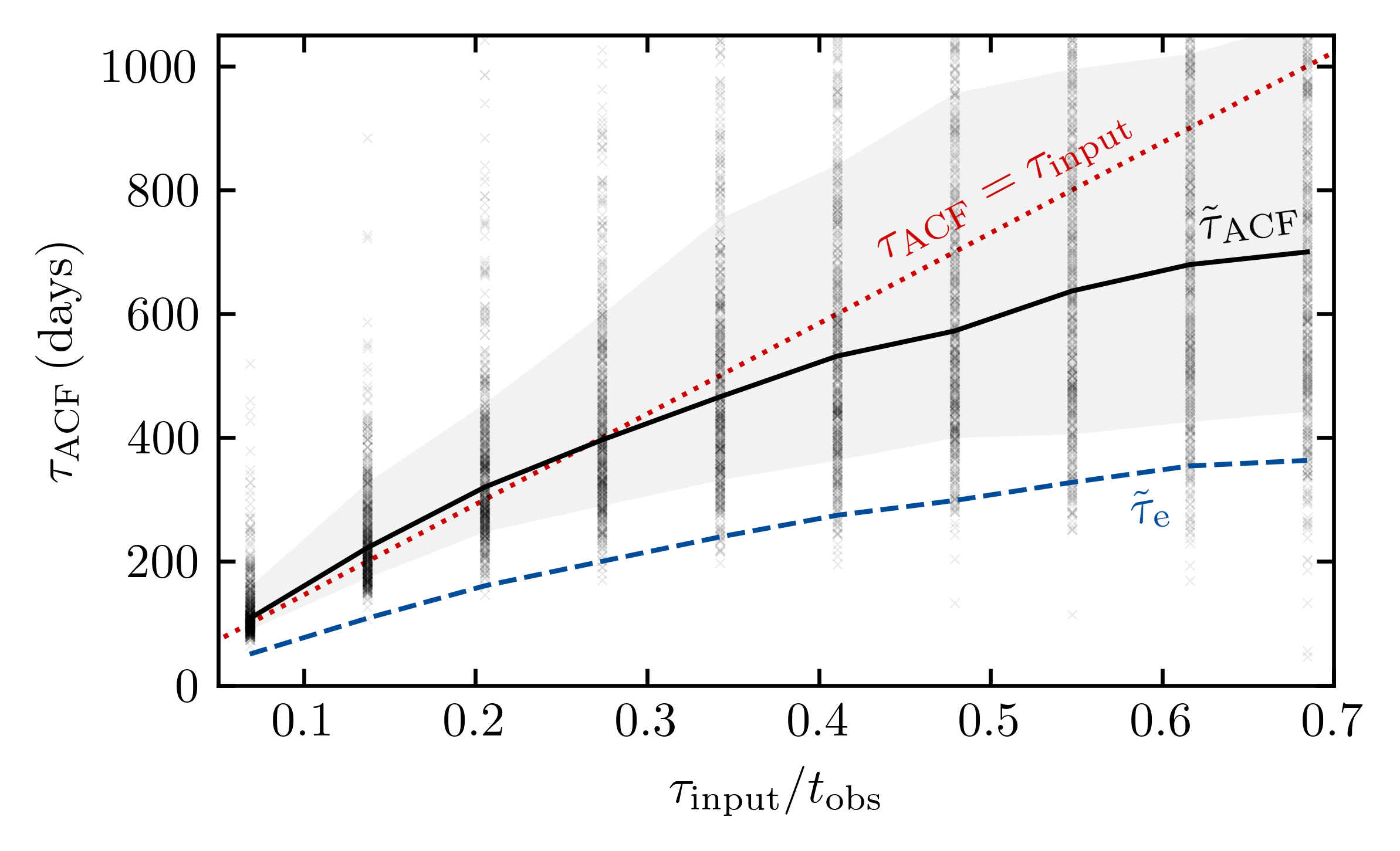}
    \caption{Autocorrelation timescale as a function of the input lifetime normalized to the observation length for the reference set of light curves. The black crosses show the results for individual simulations. Here, all spots in the same simulation have the same area, hence lifetime. The black solid line and the gray region indicate the median and the 16\textsuperscript{th} and 84\textsuperscript{th} percentiles. The red dotted line indicates \tinput. For comparison, the blue dashed line marks the median $e$-folding times, $\tilde{\tau}_e$.The fixed properties of the data are: $t_\text{obs}=4$ years; $i=70^\circ$; $\langle L\rangle=15^\circ$ and $\sigma_\text{L}=5^\circ$; $\gamma=0.2$; $P_\text{eq}=25$ days and $\alpha=0.2$.}% (500 per $\tau_\text{input}$)
    \label{fig:stdsimul}
\end{figure}

Figure~\ref{fig:stdsimul} shows the results for the reference artificial light curves. In summary (see details in Section~\ref{sec:data} and in Table~\ref{tab:parameters}), the reference data are characterized by: observation length of 4 years; stellar inclination angle of $i=70^\circ$; solar rotation profile; mean and standard deviation of spot latitudinal distribution of $15^\circ$ and $5^\circ$ respectively; and slow spot evolution. For each light curve, spots are generated randomly, all with the same maximal area $A_\text{maximal}$ and, consequently, lifetime. At a fixed input lifetime $\tau_\text{input}$ (from 100 to 1000 days with steps of 100), we obtain 500 artificial light curves, whose individual results are shown by the black crosses. The black solid line and gray region indicate the median (\avtacf) and the 16\textsuperscript{th} and 84\textsuperscript{th} percentiles of the \tacf\ distribution for each input lifetime, respectively. The blue dashed line shows the median $e$-folding time ($\tilde{\tau}_e$), which is significantly smaller than the input lifetimes. Note that \tacf\ for some simulations is longer than the input lifetimes as the spot signature is more complex than those of Figure~\ref{fig:1-spot}.

\avtacf\ closely follows \tinput\ for relatively short-lived spots, while for long-lived spots, lifetimes are severely underestimated. Nevertheless, \avtacf\ and \tinput\ are still well correlated: the Spearman correlation coefficient between \avtacf\ and \tinput\ is 1.00, while when considering the light curves individually it is $\sim0.71$. This indicates that we can use the autocorrelation function to estimate a lower limit of the characteristic active-region lifetimes. %As shown below (Section~\ref{sec:length}), this results in part from the limited observation time ($t_\text{obs}$).

Below, we explore the impact of different stellar and spots properties on the parameter \tacf\ and determine under which circumstances \tacf\ is still found to be related to the active-region lifetimes. 

\subsection{Observation length}\label{sec:length}

For the majority of the targets observed during its main mission, \kep\ collected high-precision long-term (4 years) time-series. In contrast, the time-series provided by K2 and TESS are significantly shorter \citep{Howell2014,Ricker2014}. Can active-region lifetimes still be constrained from the ACF of such short light curves?

\begin{figure}%[h]
    \centering
    \includegraphics[width=\hsize]{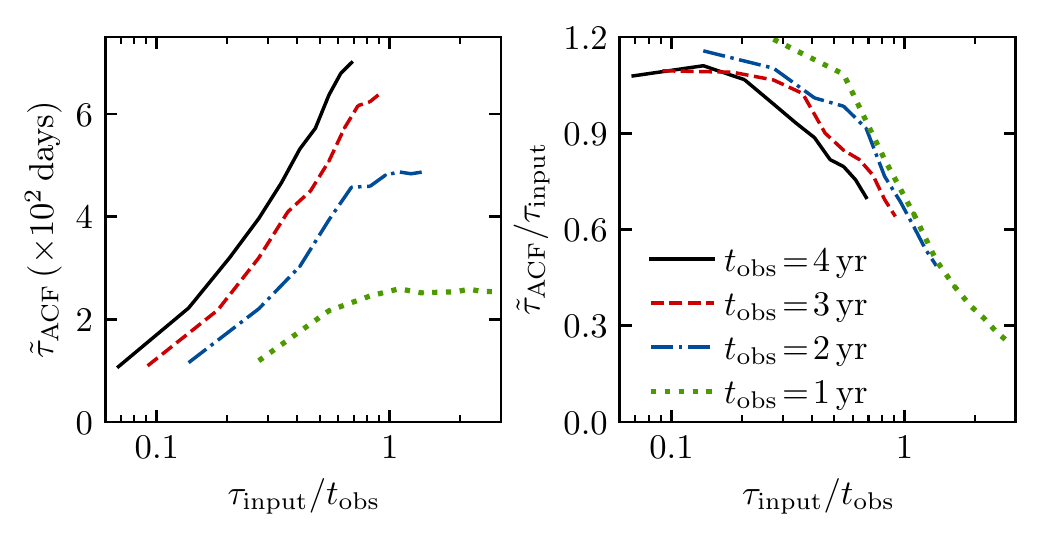}
    \caption{\avtacf\ (left) and \avtacf/$\tau_\text{input}$ (right) as a function of $\tau_\text{input}/t_\text{obs}$ for time-series of different lengths: 4 years (solid black); 3 years (dashed red); 2 years (dash-dotted blue); and 1 year (dotted green). Each curve is based on 5000 artificial light curves. }%{\it Right:} \tacf/$\tau_\text{input}$ distribution for each observation length. The dashed lines indicate the respective median values.
    \label{fig:length}
\end{figure}

Figure~\ref{fig:length} summarizes the results for different observation lengths. The solid black line in the left panel shows the same results as the dashed line in Figure~\ref{fig:stdsimul}, i.e. \avtacf\ for 4-year light curves. In the right-hand panel of Figure~\ref{fig:length}, \avtacf\ is normalized to the input lifetime. The red, blue, and green lines show the \avtacf\ for light curves with a length of 3, 2, and 1 years, respectively. For each observation length, we obtain 500 artificial light curves per $\tau_\text{input}$, which varies between 100 and 1000 days with steps of 100. Interestingly, \avtacf/$\tau_\text{input}$ for different observation lengths tends to follow the same behaviour, particularly at higher $\tau_\text{input}/t_\text{obs}$ values. When the input lifetime is shorter than $\sim30\%$ of the length of the time-series, \avtacf\ is close to $\tau_\text{input}$. When the input lifetime is longer than $\sim30\%$ of the observation length, the difference between \avtacf\ and the input value increases steeply. Our results suggest that for short light curves, we are unable to properly constrain the active-region lifetimes from the ACF, as the active regions might not be observed during their full lifetime. In particular, for 1-year light curves (observation length for targets at the TESS continuous viewing zone), the correlation between $\tau_\text{input}$ and \avtacf\ is significantly weaker in comparison with the longer $t_\text{obs}$: the Spearman correlation coefficient is 0.52 when considering \avtacf\ and 0.28 when considering the individual \tacf, in comparison to the respective correlation coefficients of 1.00 and 0.71 for 4-year light curves.

\subsection{Number of Spots}\label{sec:snumber}

\begin{figure}%[h]
    \centering
    \includegraphics[width=\hsize]{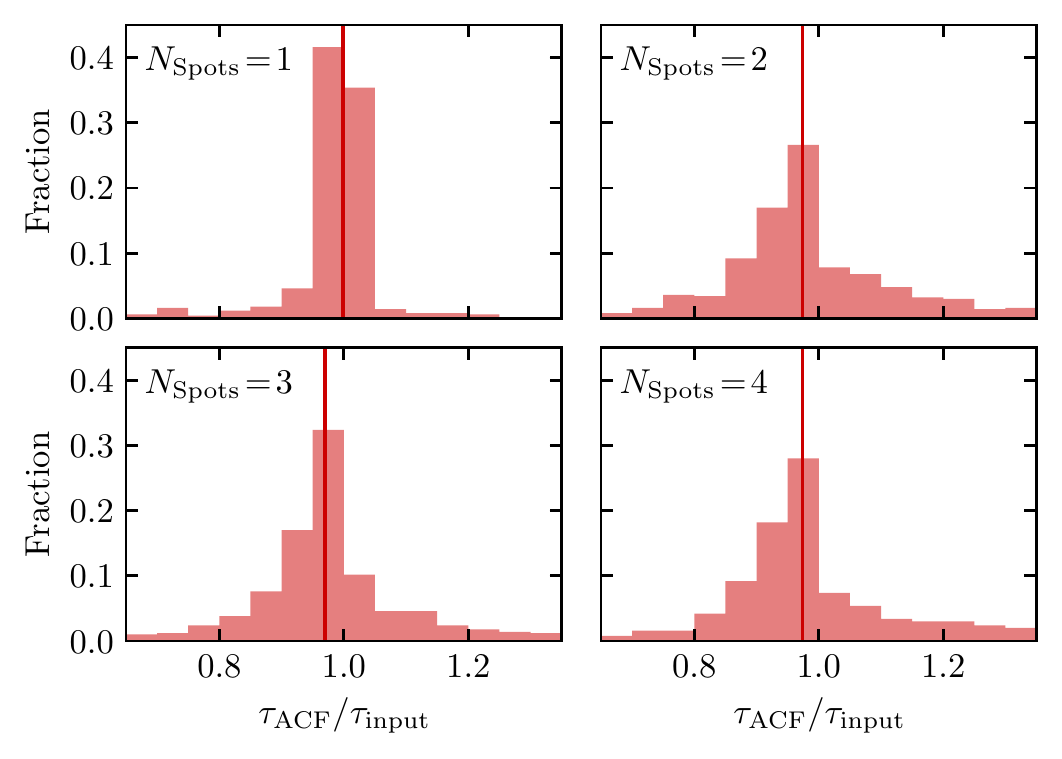}
    \caption{\tacf/$\tau_\text{input}$ distribution for fixed maximum number of spots (from 1 to 4 spots), $t_\text{obs}=4$ years, and $\tau_\text{input}=400$ days. Each panel shows the results for 500 simulations. The red lines mark the median values of each distribution.}
    \label{fig:nspots}
\end{figure}

In this exercise, we fix the maximum number of spots allowed on the stellar surface at a given time. The goal of this exercise is to determine whether the number of spots producing the signal affects the estimate \tacf. We find that the median \tacf/$\tau_\text{input}$ does not change significantly with the number of spots, varying between 0.97 and 1.00 at fixed $\tau_\text{input}/t_\text{obs}\sim0.27$ ($\tau_\text{input}=400$ days; Figure~\ref{fig:nspots}), while the scatter slightly increases as the complexity of the light curve increases.

\begin{figure}%[h]
    \centering
    \includegraphics[width=\hsize]{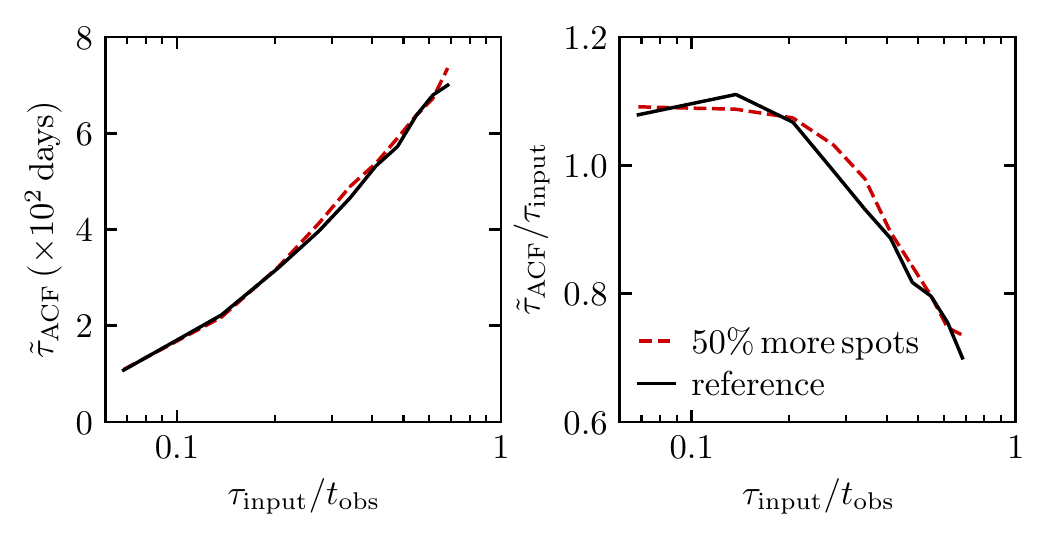}
    \caption{Same as in Figure~\ref{fig:length} but comparing the results for the reference set of light curves (black; same as dashed line in Figure~\ref{fig:stdsimul}) with those obtained when forming about $50\%$ more spots (red). The observation length is $t_\text{obs}=4$ years.}
    \label{fig:1.5xN}
\end{figure}

Next, instead of fixing the number of allowed spots for a given $\tau_\text{input}$, we increase the probability of forming spots at each time step by $50\%$. Figure~\ref{fig:1.5xN} compares \avtacf\ for the reference artificial data (same data as in Figure~\ref{fig:stdsimul}) and the distribution obtained when there are in average 1.5 times the number of spots of the reference data. The results are identical, with the reference data, i.e. with less spots, having slightly smaller \avtacf\ than the data with more spots.

%\pagebreak

\subsection{Stellar inclination}\label{sec:inc}

\begin{figure}%[h]
    \centering
    \includegraphics[width=\hsize]{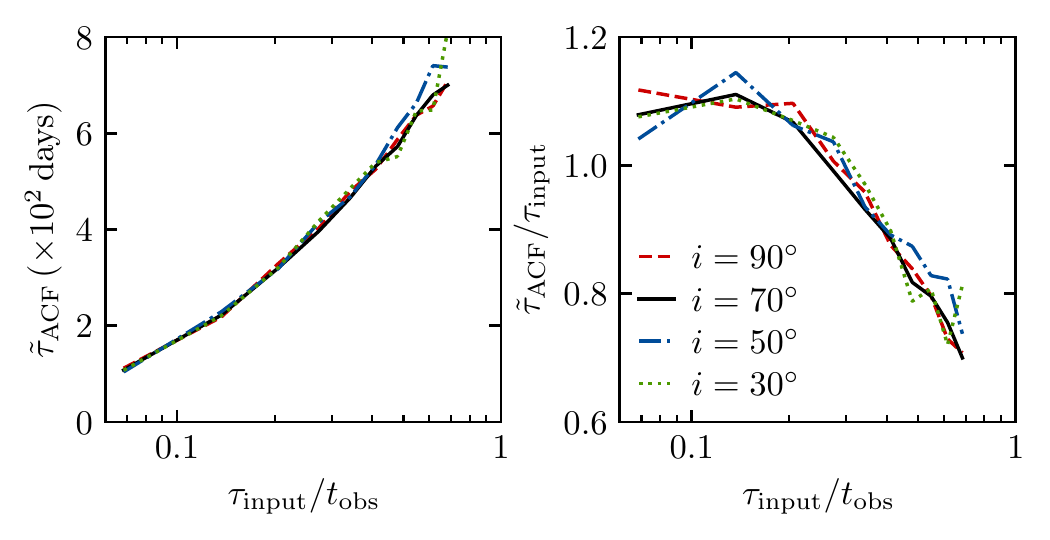}
    \caption{Same as in Figure~\ref{fig:length} but for different inclination angles ($i=90^\circ$ - red; $i=70^\circ$ - black; $i=50^\circ$ - blue; $i=30^\circ$ - green). The observation length is $t_\text{obs}=4$ years.}
    \label{fig:inc}
\end{figure}

The stellar inclination angle affects the spot visibility and, consequently, the spot modulation of light curves. But can it affect the spot properties we may be able to constrain? Here, we vary the stellar inclination angle, while the remainder of the parameters are kept unchanged. We find that the stellar inclination does not affect the decay timescale of the ACF and consequently \tacf\ (Figure~\ref{fig:inc}). As long as spots cross the visible disc of the star and spot modulation is detected, \tacf\ can be measured, providing a lower limit to the active-region lifetimes particularly for long \tinput.

\subsection{Active longitudes}\label{sec:longitudes}

\begin{figure}%[h]
    \centering
    \includegraphics[width=\hsize]{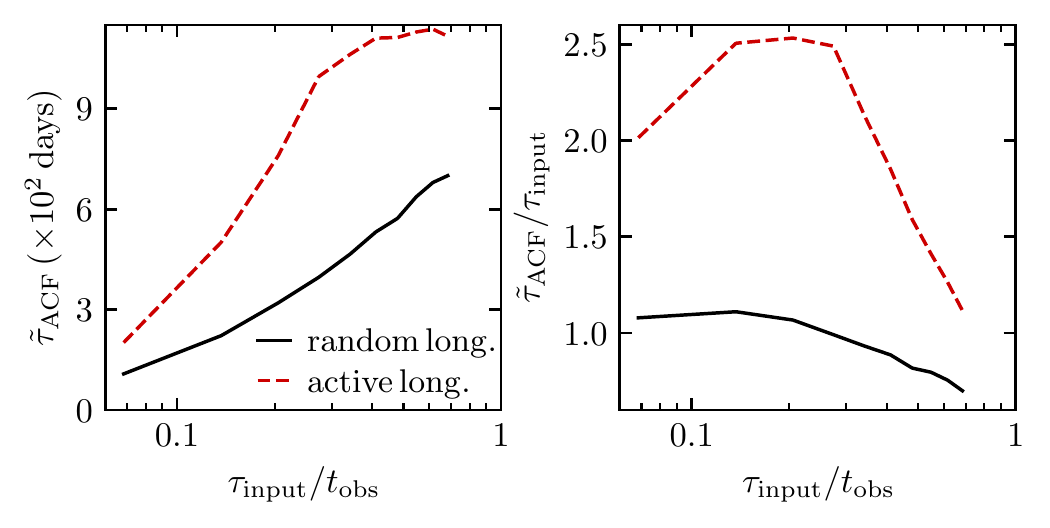}
    \caption{Same as in Figure~\ref{fig:length} but for random (black) and active (red) spot longitudes. The observation length is $t_\text{obs}=4$ years.}
    \label{fig:activelong}
\end{figure}

For the reference light curves, spots emerge at random longitudes (see Sec.~\ref{sec:data}). However, spots tend to be formed in active regions \citep[e.g.][]{Bumba1965,Bogart1982,Mathur2014,Lanza2019}. Active regions live longer than individual spots. For the set of light curves represented in red in Figure~\ref{fig:activelong} we force the spots to emerge consecutively at nearly the same longitude. In this case, one would expect that \tacf\ is related to the active-region lifetime given that the spots' signature would have consecutively the same phase. Indeed, Figure~\ref{fig:activelong} shows that, while $\tau_\text{input}$ is still the lifetime of individual spots, when spots are forced to emerge in active longitudes (red) the average \tacf\ increases significantly in comparison with the case of random positions (black). Our results suggest that when spots form in active regions \tacf\ is related to the active-region lifetime, while when spots form individually \tacf\ is related to the spot lifetimes. Note that as found in Section~\ref{sec:length}, \tacf\ is severely affected by the observation time. In particular, for the light curves obtained with active longitudes with $\tau_\text{input}\gtrsim0.2t_\text{obs}$, \avtacf\ is mostly constant independently on the adopted \tinput.

\subsection{Surface rotation profile and spot formation zone}\label{sec:rot}

In this section, by varying the rotation profile and the spot formation zone separately, we investigate how the range of rotation rates probed by the spots impacts \tacf.

\begin{figure}
    \centering
    \includegraphics[width=\hsize]{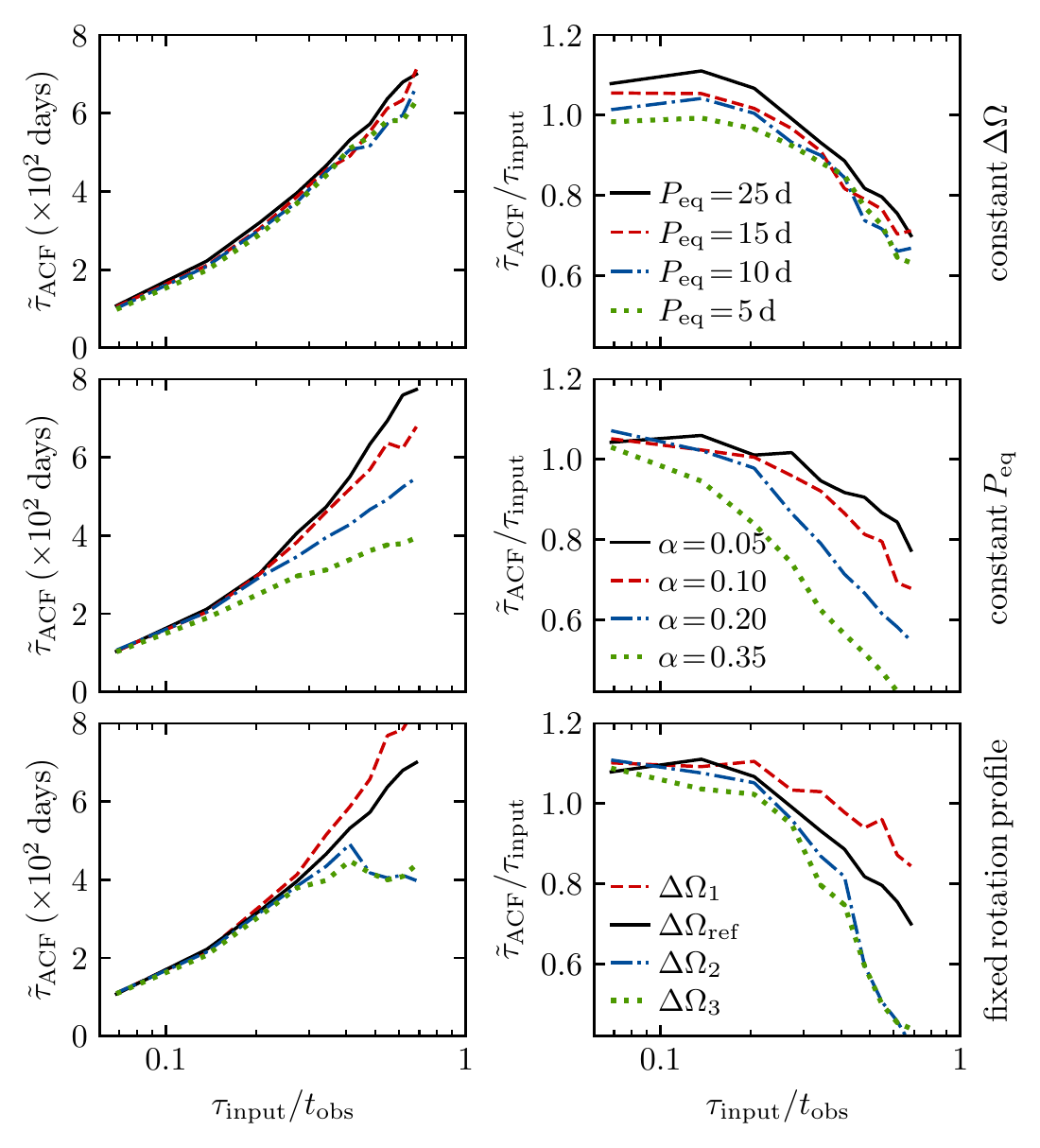}
    \caption{Same as in Figure~\ref{fig:length} but for different rotation profiles and spot formation zones. {\it Top:} the Differential rotation is fixed at solar value, while the rotation period at the equator is varied ($P_\text{eq}=25$ days - solid black; $P_\text{eq}=15$ days - dashed red; $P_\text{eq}=10$ days - dash-dotted blue; and $P_\text{eq}=5$ days - dotted green). {\it Middle:} $P_\text{eq}$ is fixed at 15 days, while differential rotation is varied ($\alpha=0.05$ - solid black; $\alpha=0.10$ - dashed red; $\alpha=0.20$ - dash-dotted blue; and $\alpha=0.35$ - dotted green). {\it Bottom:} The rotation profile (both $P_\text{eq}$ and $\alpha$ are fixed, while the spot-latitude distribution is varied ($\langle L\rangle=5^\circ, \sigma_\text{L}=5^\circ$, $\Delta\Omega_1$ - dashed red; $\langle L\rangle=15^\circ, \sigma_\text{L}=5^\circ$, $\Delta\Omega_\text{ref}$ - solid black; $\langle L\rangle=25^\circ, \sigma_\text{L}=5^\circ$, $\Delta\Omega_2$ - dash-dotted blue; $\langle L\rangle=15^\circ, \sigma_\text{L}=10^\circ$, $\Delta\Omega_3$ - dotted green). The observation length is $t_\text{obs}=4$ years.}
    \label{fig:rot}
\end{figure}

For all the light curves above we adopted a nearly solar rotation profile (see Section~\ref{sec:data}). The reference data set is shown by the black line in the top and bottom panels of Figure~\ref{fig:rot} (same as shown by the dashed line in Figure~\ref{fig:stdsimul}). In the top panels, the latitudinal differential rotation $\Delta\Omega$ is fixed matching that of the black line, while the rotation period at the equator $P_\text{eq}$ is varied. Similarly to the data sets above, each curve represents \avtacf/\tinput\ for 500 realizations per \tinput\ (from 100 to 1000 days with steps of 100). The curves for the different rotation profiles tend to overlap. This indicates that \avtacf\ is independent of the average rotation rate, if the differential rotation is unchanged. In the middle row, $P_\text{eq}$ is fixed at 15 days and the differential rotation is changed by varying the parameter $\alpha$. The stronger the differential rotation is, the larger is the difference between \avtacf\ and $\tau_\text{input}$, with the spot/active-region lifetime being significantly underestimated. This results from the fact that strong differential rotation yields to fast beating patterns in the light curve, particularly if spots are long-lived leading to stable beating signals. Both stronger differential rotation and long-lived spots are observed in fast rotating and very active stars \citep[e.g.][]{Strassmeier2002,Reinhold2013a}. When a stable beating pattern is present in a light curve, the ACF is also affected showing itself fast beating, which in turn affects the ACF decay. In the bottom panels of Figure~\ref{fig:rot}, the rotation profile is fixed at the solar rotation, but the spot formation zone is changed. In a differentially rotating star, this means that the spots responsible for the rotation signal probe different ranges of rotation rates. Thus, the width of the spot formation zone is expected to affect \tacf. In particular, considering a wider spot formation zone would be similar to considering stronger differential rotation. This is shown in the bottom panels of Figure~\ref{fig:rot}. The wider the range of rotation rates of the spots, the more underestimated the spot lifetimes are. The legend of the bottom panels is ordered from narrowest (top; dashed red) to widest (bottom; dotted green) range of rotation rates, where ``ref" stands for reference data set.

\subsection{Spot growth and decay}\label{sec:evolution}

By affecting the coherence of the signal, spot evolution, i.e. change in spot area, is expected to affect the autocorrelation function of the light curve and, hence, \tacf. The growth and decay rates can be described through a power law. For the artificial data above, the growth and decay rates are fixed, with $\gamma=0.2$. In this section, we assess the impact of spot evolution on the parameter \tacf\ by varying the exponent $\gamma$. For simplification purposes, the growth and decay rate are considered to be equal. 

\begin{figure}%[h]
    \centering
    \includegraphics[width=\hsize]{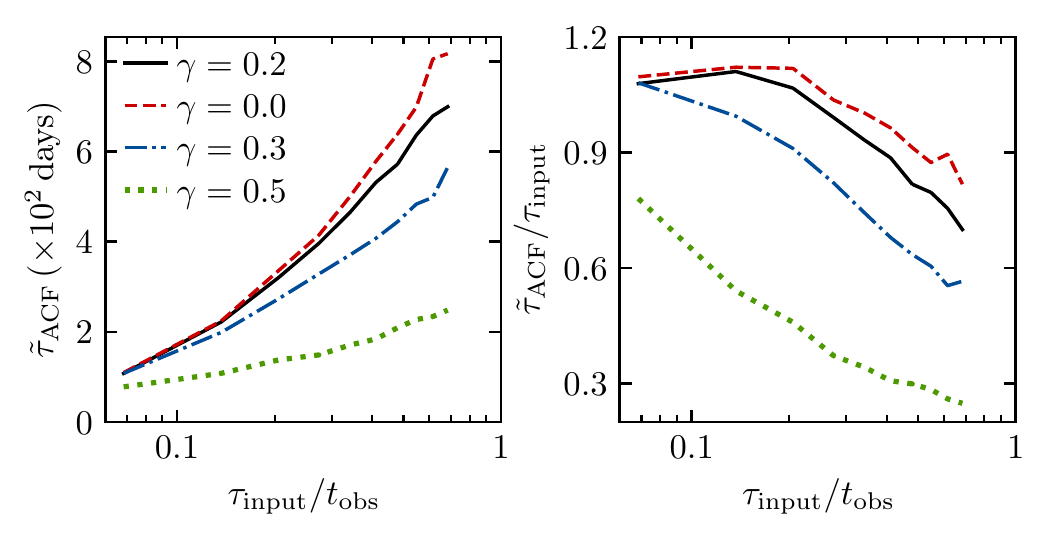}
    \caption{Same as in Figure~\ref{fig:length} but for different spot evolution rates: $\gamma=0.0$ represents the set of artificial data with no spot evolution, while $\gamma=0.5$ corresponds to the case of fastest spot evolution ($\gamma=0.0$ - dashed red; $\gamma=0.2$ - solid black; $\gamma=0.3$ - dash-dotted blue; $\gamma=0.5$ - dotted green). The observation length is $t_\text{obs}=4$ years.}
    \label{fig:evol}
\end{figure}

Figure~\ref{fig:evol} compares the results for the reference set of artificial data (black; same as the dashed line in Figure~\ref{fig:stdsimul}) and: the case of no spot evolution; and two cases of fast spot growth and decay. Spot evolution is found to significantly affect the value for the observed spot timescale. The faster the spot evolution, the more underestimated the active-region lifetime is. These results reiterate that \tacf\ is a lower limit of the characteristic active-region lifetimes. 

\subsection{Random spot lifetimes}\label{sec:vA}

\begin{figure}
    \centering
    \includegraphics[width=\hsize]{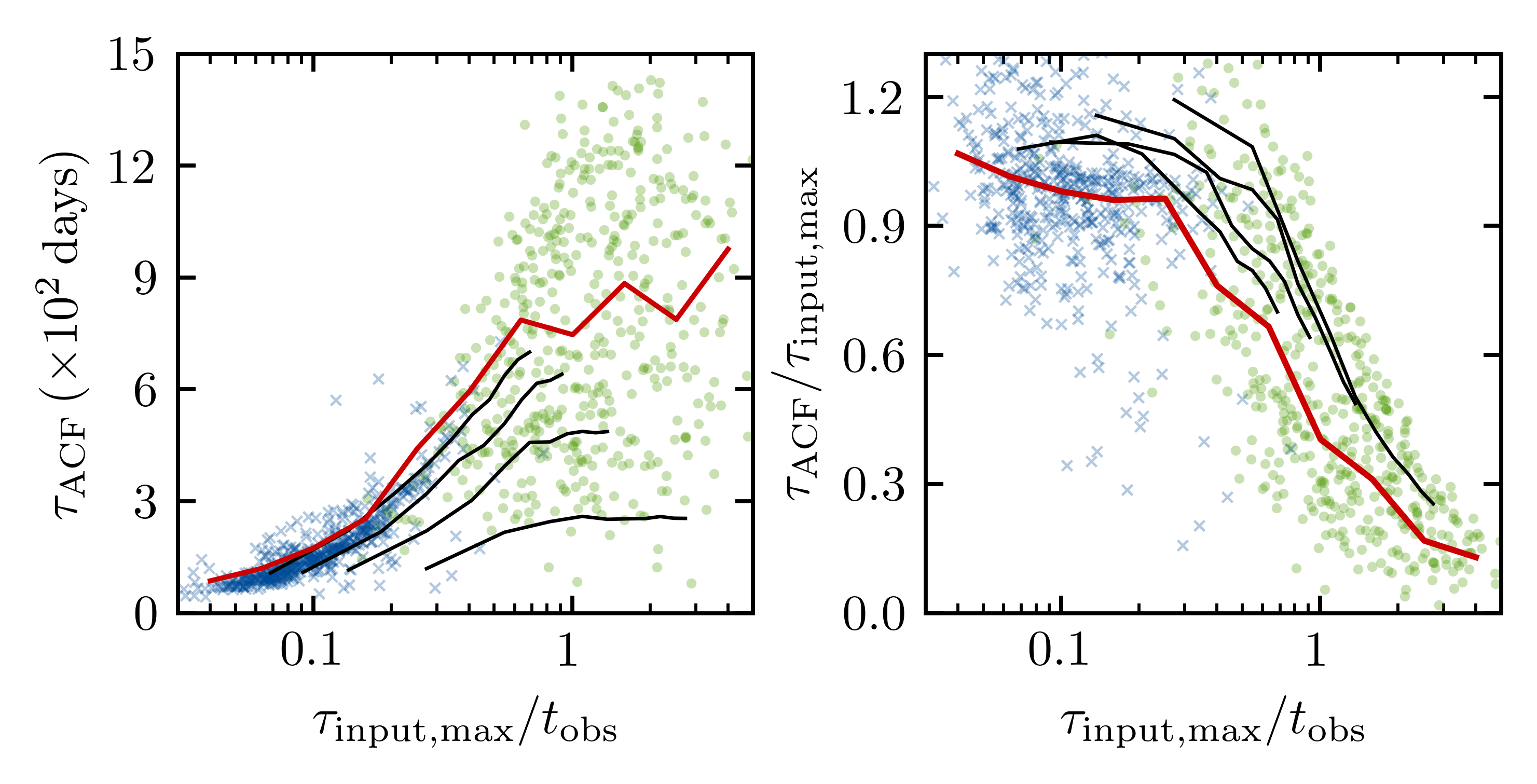}
    \caption{\tacf\ and \tacf/$\tau_\text{input}$ as a function of $\tau_{\text{input},\text{max}}/t_\text{obs}$, where $\tau_{\text{input},\text{max}}$ is the lifetime of the largest, longest-lived spot in each light curve. In this exercise, the maximal area of each spot is random according to a log-normal distribution. Similarly to the previous data sets, we adopt the Gnevyshev-Waldmeier rule, thus lifetimes are proportional to the spot areas. Blue and green symbols show the results for 1000 individual light curves. For the data set represented in blue, the distribution, from which spot areas are randomly drawn, peaks at $\sim150\mu\text{Hem}$, while for the data set in green the distribution peaks at $\sim2000\mu\text{Hem}$. The solid red lines mark the median \tacf/$\tau_{\text{input},\text{max}}$. $t_\text{obs}$ in this exercise (symbols and red line) is fixed at 4 years. For reference, the solid black lines show the median \tacf/$\tau_\text{input}$ for the artificial data in Figure~\ref{fig:length} where different $t_\text{obs}$ where considered.}\label{fig:vA}
\end{figure}

For the artificial data above, $\tau_\text{input}$ was fixed within the same simulation with all spots having the same $A_\text{maximal}$ and, thus, lifetime. In this section, the spot areas are random, i.e. within the same simulation different spots have different areas and lifetimes. As above, the lifetimes are proportional to the spot areas according to the Gnevyshev-Waldmeier rule (see Section~\ref{sec:data}). Figure \ref{fig:vA} shows the results for two data sets of 500 light curves each. For the first data set (blue crosses), the maximal spot areas are drawn from a log-normal distribution with mean and standard deviation of 5.8 and 1, respectively, which peaks around 150 $\mu\text{Hem}$, i.e. the spot lifetime distribution peaks around 15 days. For the second data set (green circles), the mean and standard deviation of the log-normal distribution is 8.5 and 1, respectively, with the distribution peaking around 2000 $\mu\text{Hem}$, corresponding to a lifetime of 200 days. Note that the maximal area distribution for the Sun is consistent with a log-normal distribution \citep[e.g.][see also \citet{Santos2015}]{Bogdan1988,Baumann2005,Hathaway2008}.

In Figure~\ref{fig:vA}, \tacf\ is normalized by the lifetime of the largest spot (maximum lifetime; $\tau_{\text{input},\text{max}}$) in the simulation. The red solid line shows the normalized \avtacf\ for the 1000 light curves (blue and green symbols). For reference, we show in black the results from Figure~\ref{fig:length} obtained with different $t_\text{obs}$. Note however that $t_\text{obs}$ for the blue and green symbols and, consequently, for the red line, is fixed at 4 years, whereas $\tau_{\text{input},\text{max}}$ for some of the light curves is longer than those in Figure~\ref{fig:length}.  For this exercise, while \tacf\ values are relatively smaller, they tend to show a similar behaviour to the results shown in black, which reinforces the assessment of \tacf\ being limited by the length of the observations (Section~\ref{sec:length}). Particularly, at large input lifetimes, \tacf\ is mostly independent of the input lifetime. We adopt $\tau_{\text{input},\text{max}}$ because depending on the latitudinal distribution of spots and stellar inclination, the largest, long-lived spots usually dominate the rotational signal. Also, \tacf\ for the relatively short-lived spots tends to overestimate both mean and median \tinput\ values.

%\pagebreak

\section{Conclusions}\label{sec:conclusion}

The autocorrelation function of stellar light curves has been extensively used in the literature to constrain surface rotation periods \citep[e.g.][]{McQuillan2013a,McQuillan2014,Garcia2014,Santos2019a,Santos2021}. In addition, as the ACF reflects the coherence of the signal, the decay timescale of the ACF is expected to be related to the spot or active-region lifetimes \citep[e.g.][]{Lanza2014}. Subsequently, the ACF was used by \citet{Giles2017} to constrain the active-region lifetimes for about 2,200 solar-type stars observed by {\it Kepler}. The authors adopted an underdamped harmonic oscillator, characterized by an exponential decay, to model the ACF.

The main goal of this work is to determine whether the decay timescale of the ACF can be used to estimate the spot/active-region lifetimes and determine under which conditions the lifetimes can still be constrained. Using the tools developed in \citet{Santos2015,Santos2017a}, we obtained artificial light curves with different observation, stellar, and spot properties.

The first data set comprises one-spot light curves. We initially modelled the respective ACFs using an underdamped harmonic oscillator. Even for such simple signals, we found that the retrieved $e$-folding time is only about half of the input lifetimes.

Upon a thorough inspection of the ACF and its amplitude decay as a function of the temporal lag, we concluded that an exponential decay is not the most appropriate function to describe the ACF. In fact, the ACF decay is linear for simple unperturbed signals. For the one-spot light curves, in particular, the recovered timescale for the linear decay is approximately the input spot lifetime, while the $e$-folding time was only about half of the input value.

Therefore, for the remainder of the analysis, we proposed and used a linear decay to model the ACF of the light curves, where the ACF timescale, \tacf, corresponds to the observed spot/active-region timescale. For an unperturbed simple spot modulation of the light curve, \tacf\ matches the input lifetime.

We then carried out a series of control tests by varying separately a number of observation, spot, and stellar properties in order to assess how each of them affects the constraint on spot/active-region lifetimes from the ACF. In the first exercises, all spots in a given simulation have the same lifetime, while in the last exercise spots have different lifetimes.

We found that the inferred timescale \tacf is greatly restricted by the observation length of the light curves. For the lifetimes shorter than 1/3 of the observation length, \avtacf\ is closely related to the input lifetimes, while for longer lifetimes \avtacf\ underestimates \tinput. Thus, our results indicate that \tacf\ is a lower limit of the characteristic spot/active-region lifetimes. Particularly, for 1-year light curves \tacf\ and \tinput\ are mostly uncorrelated. The underestimation of \tinput\ is the most significant for long-lived spots or active regions. 

%We find that the number of spots and stellar inclination do not affect significantly the autocorrelation timescale. 

The decay of the ACF is also significantly affected by differential rotation and spot evolution. The stronger the differential rotation or the wider the spot formation zone, i.e. the wider the range of rotation rates associated to spots, the more severely underestimated the spot/active-region lifetimes are. Fast spot evolution also leads to shorter inferred timescales in comparison with slow spot evolution. Nevertheless, \tacf\ was still found to be related to the input lifetime. As follows, \tacf\ is still a valid lower limit.

We found that the effect from the inclination angle, number of spots, and average rotation rate, while keeping the remainder of the properties fixed, is negligible.

We finally consider spots of different sizes and, thus, different lifetimes in the same light curve. Under these circumstances, \tacf\ is still linked to the input lifetimes. We took the lifetime of the longest-lived spot, i.e. the maximum lifetime, in a given simulation as the reference input lifetime. \tacf\ is typically shorter than $\tau_{\text{input},\text{max}}$, but is longer than the average input lifetime. Furthermore, similarly to the previous control tests, \tacf\ is significantly impacted by the observation length. Particularly, for spot/active-region lifetimes that are longer than about 1/3 of the length of the light curve, there is a sharp increase of the difference between \tacf\ and the input lifetime. Nevertheless, for lifetimes shorter than the length of the light curve, \tacf\ can still be used as a lower limit to the true lifetime.

In summary, our results indicate that \tacf\ underestimates the characteristic spot/active-region lifetimes, being a lower limit to the true lifetimes.
For real data, in order to prevent substantial underestimation of the active-region lifetimes, it might be important to avoid or flag light curves with signatures of rapidly evolving spots and fast beating patterns. While spot evolution may be hard to tackle, beating patterns can be identified through photometric magnetic activity metrics like \sph\ \citep{Mathur2014}, as beating patterns affect such metrics. One group of targets, in particular, flagged in \citet{Santos2019a,Santos2021} as close-in binary candidates, exhibits stable and, often, fast beating patterns. These targets have similar behaviour to targets identified as tidally-synchronized binaries by \citet[][see further discussion in \citealt{Santos2019a,Santos2021}]{Simonian2019}. Beating patterns are also a source of concern in other types of analyses, for example, leading to positives in the activity-cycle search \citep[e.g. see discussion in][]{Mathur2014}.

More critical to the observed active-region timescale than the effect from beating, the length of the time-series plays an important role on our ability to constrain active-region lifetimes. To date, the long-term photometry of the \kep\ main mission still constitutes the best data set to measure active-region lifetimes. However, one still should keep in mind that for very long-lived active regions (for example those reported in super-flaring stars), even the four years of \kep\ data might be insufficient. In Santos et al. (in preparation), we will investigate the autocorrelation timescale for \kep\ targets with known rotation periods from \citet{Santos2019a,Santos2021}. 
In the future, long-term observations from possible TESS extended missions and from the forthcoming PLATO mission \citep{Rauer2014} may also provide suitable data for such studies.

Finally, we note that the results from this work do not invalidate those in \citet{Giles2017}. The timescale retrieved in their work is still a lower limit of the active-region lifetimes. In particular, as described above the $e$-folding time is only about half of the true lifetimes even for simple one-spot light curves. The trends between the retrieved timescale and effective temperature and the amplitude of the rotational modulation found by the authors is still correct. Further discussion will be presented in Santos et al. (in preparation).

\begin{table*}\centering
\begin{tabular}{cll}
\hline
Parameter & Meaning & Reference data set\\\hline
$t_\text{obs}$ & Observation length & 4 yrs\\
$i$ & Stellar inclination angle & $70^\circ$\\
$\langle L\rangle$ & Average spot latitude & $15^\circ$\\
$\sigma_\text{L}$ & Standard deviation of the spot latitudinal distribution & $5^\circ$\\
$A_\text{maximal}$ & Maximal area of a given spot & $1000\leq A_\text{maximal}\leq 10000\,\mu \text{Hem}$\\
\tinput & Input spot lifetime: $\tau_\text{input}=A_\text{maximal}/D_\text{GW}$ & $100\leq\tau_\text{input}\leq 1000$ days\\
$D_\text{GW}$ & Constant of proportionality in the Gnevyshev-Waldmeier rule & $10\,\mu\text{Hem\,day}^{-1}$\\
$\gamma$ & Spot evolution rate exponent & 0.2\\
$P_\text{eq}$ & Surface rotation period at the equator & 25 days\\
$\alpha$ & Surface differential rotation shear & 0.2\\
$\Delta\Omega$ & Surface differential rotation &\\
$\tau_{\text{input},\text{max}}$ & Input lifetime of the longest-lived spot when considering random lifetimes & \\
$t_\text{vis}$ & Spot visibility time: time that the spot is visible during a single rotation &\\
$\tau_e$ & $e$-folding time of the ACF (exponential decay) &\\
$\tilde{\tau}_e$ & Median $e$-folding time of the ACF &\\
\tacf & Observed spot/active-region timescale &\\
\avtacf & Median observed spot/active-region timescale& \\\hline
\end{tabular}
\caption{Summary of the parameters in this study. The last column indicates the values adopted in the reference data set shown in Figure~\ref{fig:stdsimul}.}\label{tab:parameters}
\end{table*}

\section*{Acknowledgements}

We thank the referee Prof. Suzanne Aigrain for the constructive comments that helped to improve the manuscript.
The material is supported by the National Aeronautics and Space Administration (NASA) under Grant No. NNX17AF27 to the Space Science Institute (Boulder, CO USA). ARGS acknowledges the support STFC consolidated grant ST/T000252/1. SM acknowledges support by the Spanish Ministry of Science and Innovation with the Ramon y Cajal fellowship number RYC-2015-17697 and the grant number PID2019-107187GB-I00. RAG acknowledges the support from PLATO and GOLF CNES grants. MSC and PPA acknowledge  FCT/MCTES for support through the research grants UIDB/04434/2020, UIDP/04434/2020 and PTDC/FIS-AST/30389/2017, and FEDER - Fundo Europeu de Desenvolvimento Regional through COMPETE2020 - Programa Operacional Competitividade e Internacionaliza\c{c}\~ao (grant: POCI-01-0145-FEDER-030389). MSC is supported by national funds through FCT in the form of a work contract.

{\it Software:} NumPy \citep{2020NumPy-Array}, SciPy \citep{2020SciPy-NMeth}, Matplotlib \citep{matplotlib}, \texttt{emcee} \citep{Foreman-Mackey2013}.

%%%%%%%%%%%%%%%%%%%%%%%%%%%%%%%%%%%%%%%%%%%%%%%%%%
\section*{Data Availability}
 
The artificial data underlying this article will be shared on reasonable request to the corresponding author.



\bibliographystyle{mnras}
\bibliography{simul,python} % if your bibtex file is called example.bib

\appendix

\pagebreak

\section{Complementary figures}

\begin{figure}
    \centering
    \includegraphics[width=\hsize]{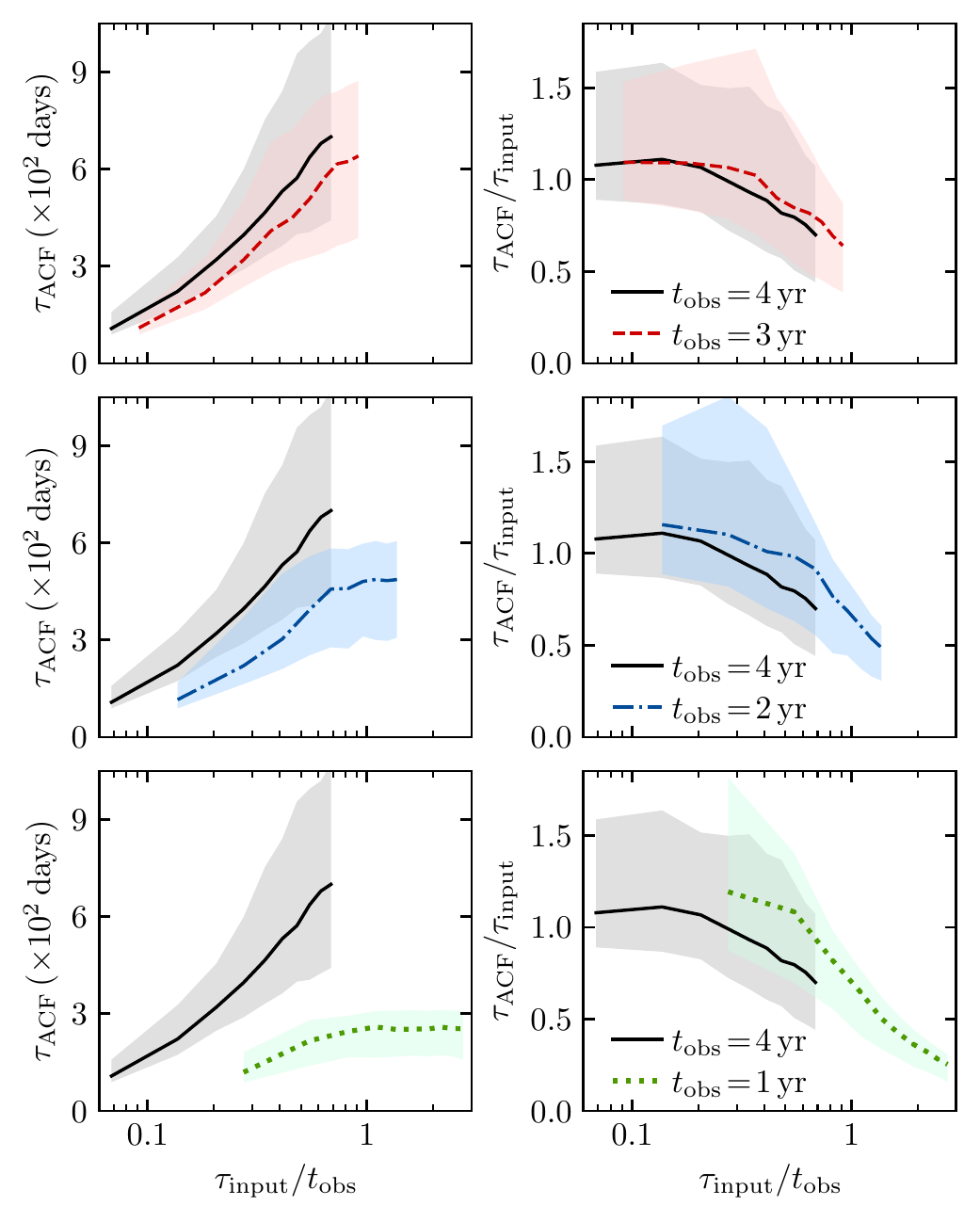}\vspace{-0.65cm}
    \caption{Same as in Figure~\ref{fig:length}, but where the shaded region indicates the 16\textsuperscript{th} and 84\textsuperscript{th} percentiles of the \tacf\ distribution for each \tinput. The solid line indicates the median \tacf\ (\avtacf).}\vspace{-0.5cm}
    \label{fig:lengthA}
\end{figure}

\begin{figure}
    \centering
    \includegraphics[width=\hsize]{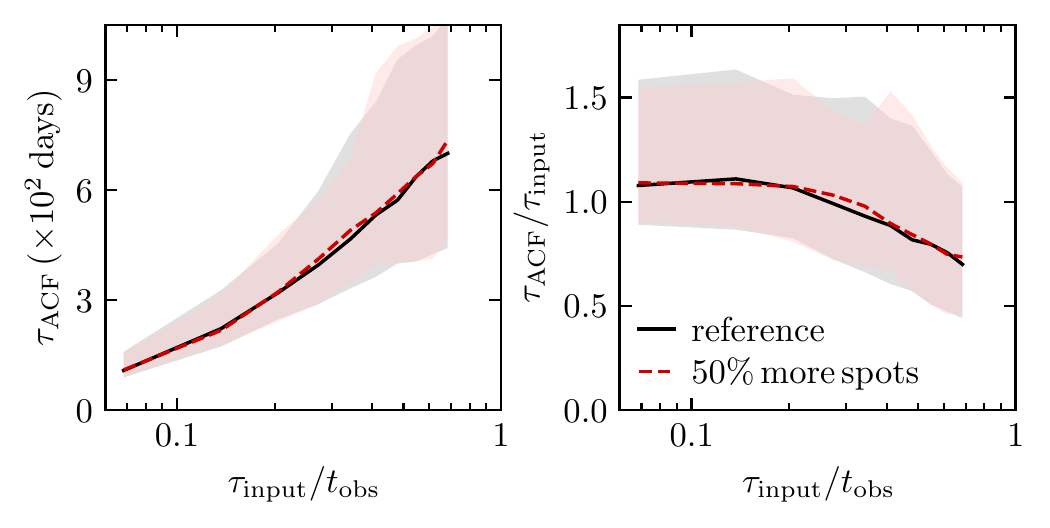}\vspace{-0.6cm}
    \caption{Same as in Figure~\ref{fig:1.5xN}, but where the shaded region indicates the 16\textsuperscript{th} and 84\textsuperscript{th} percentiles of the \tacf\ distribution for each \tinput. The solid line indicates the median \tacf\ (\avtacf).}
    \label{fig:1.5xNA}\vspace{-0.38cm}
\end{figure}

\begin{figure}
    \centering
    \includegraphics[width=\hsize]{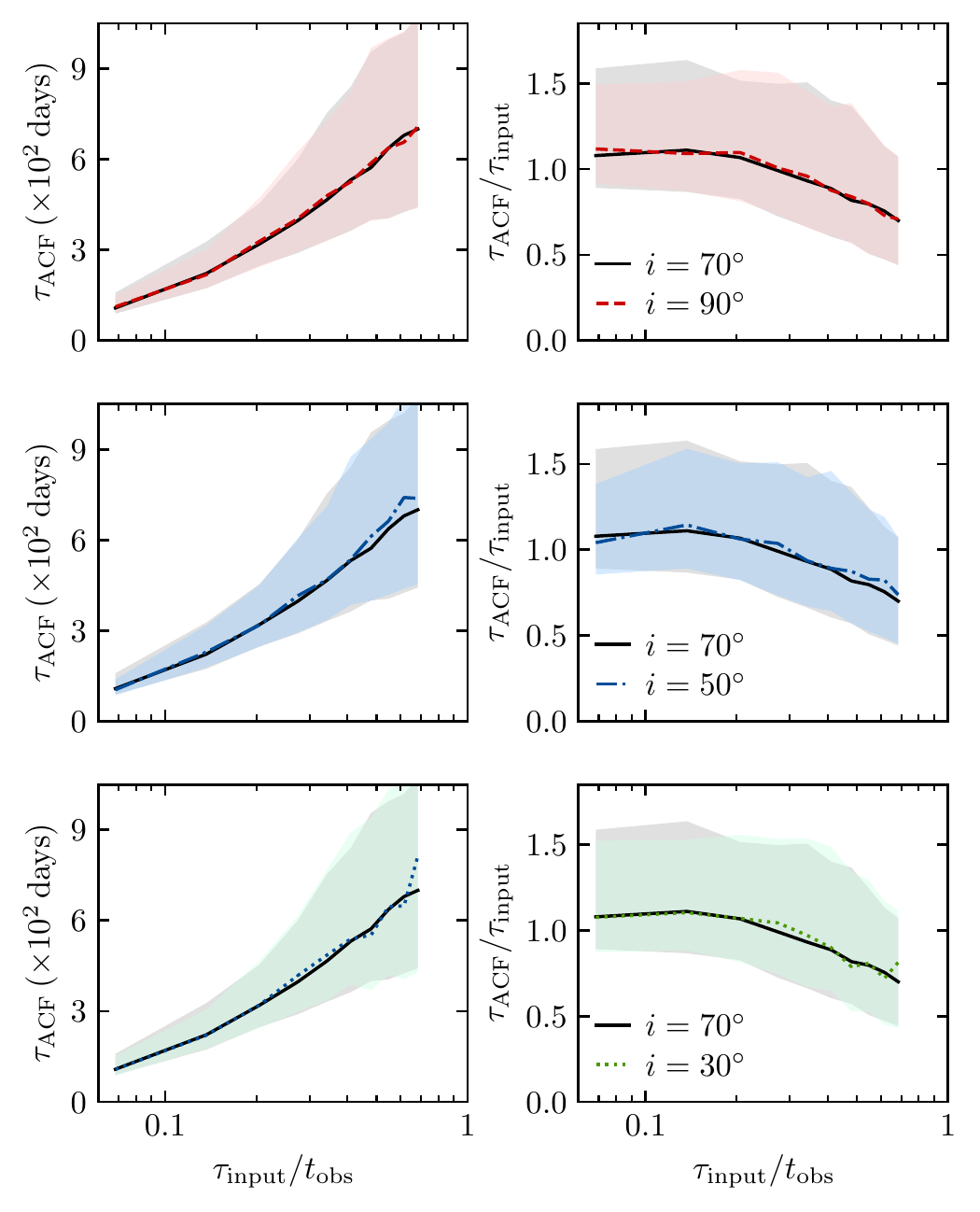}\vspace{-0.6cm}
    \caption{Same as in Figure~\ref{fig:inc}, but where the shaded region indicates the 16\textsuperscript{th} and 84\textsuperscript{th} percentiles of the \tacf\ distribution for each \tinput. The solid line indicates the median \tacf\ (\avtacf).}
    \label{fig:incA}\vspace{-0.2cm}
\end{figure}

\begin{figure}
    \centering
    \includegraphics[width=\hsize]{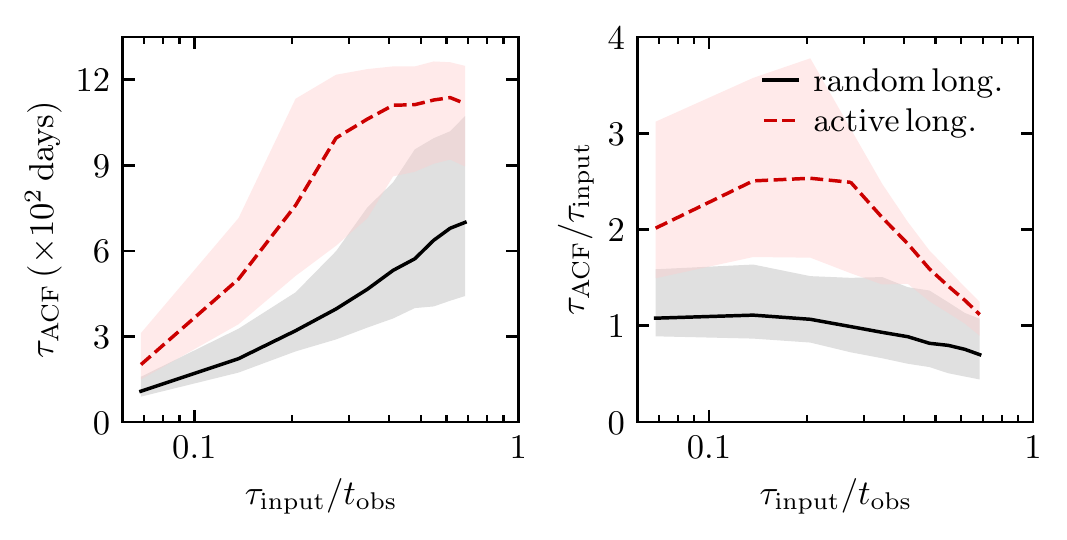}\vspace{-0.6cm}
    \caption{Same as in Figure~\ref{fig:activelong}, but where the shaded region indicates the 16\textsuperscript{th} and 84\textsuperscript{th} percentiles of the \tacf\ distribution for each \tinput. The solid line indicates the median \tacf\ (\avtacf).}
    \label{fig:activelongA}\vspace{-0.2cm}
\end{figure}

\begin{figure}
    \centering
    \includegraphics[width=\hsize]{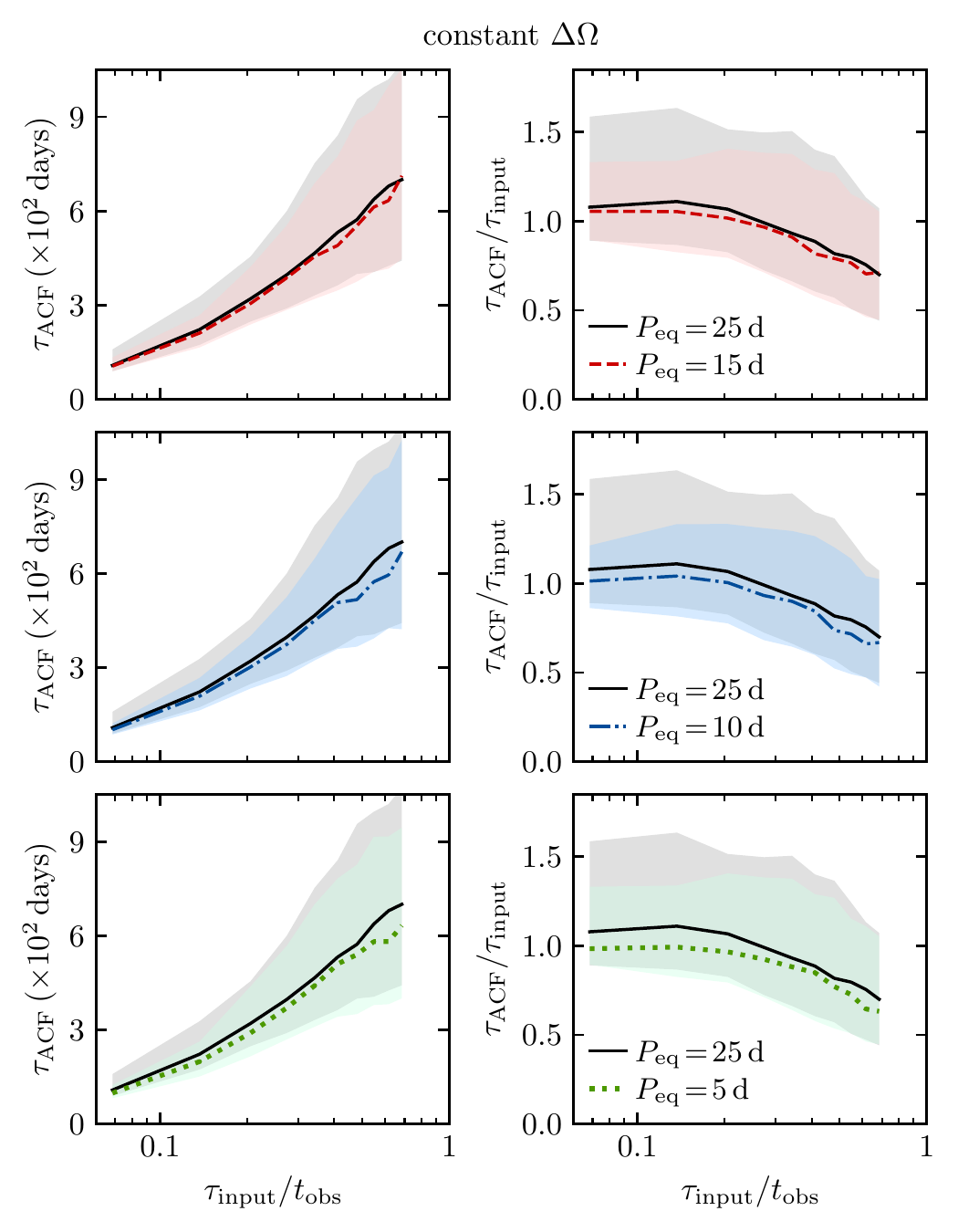}
    \caption{Same as in top panel of Figure~\ref{fig:rot}, but where the shaded region indicates the 16\textsuperscript{th} and 84\textsuperscript{th} percentiles of the \tacf\ distribution for each \tinput. The solid line indicates the median \tacf\ (\avtacf).}
    \label{fig:rot1A}
\end{figure}

\begin{figure}
    \centering
    \includegraphics[width=\hsize]{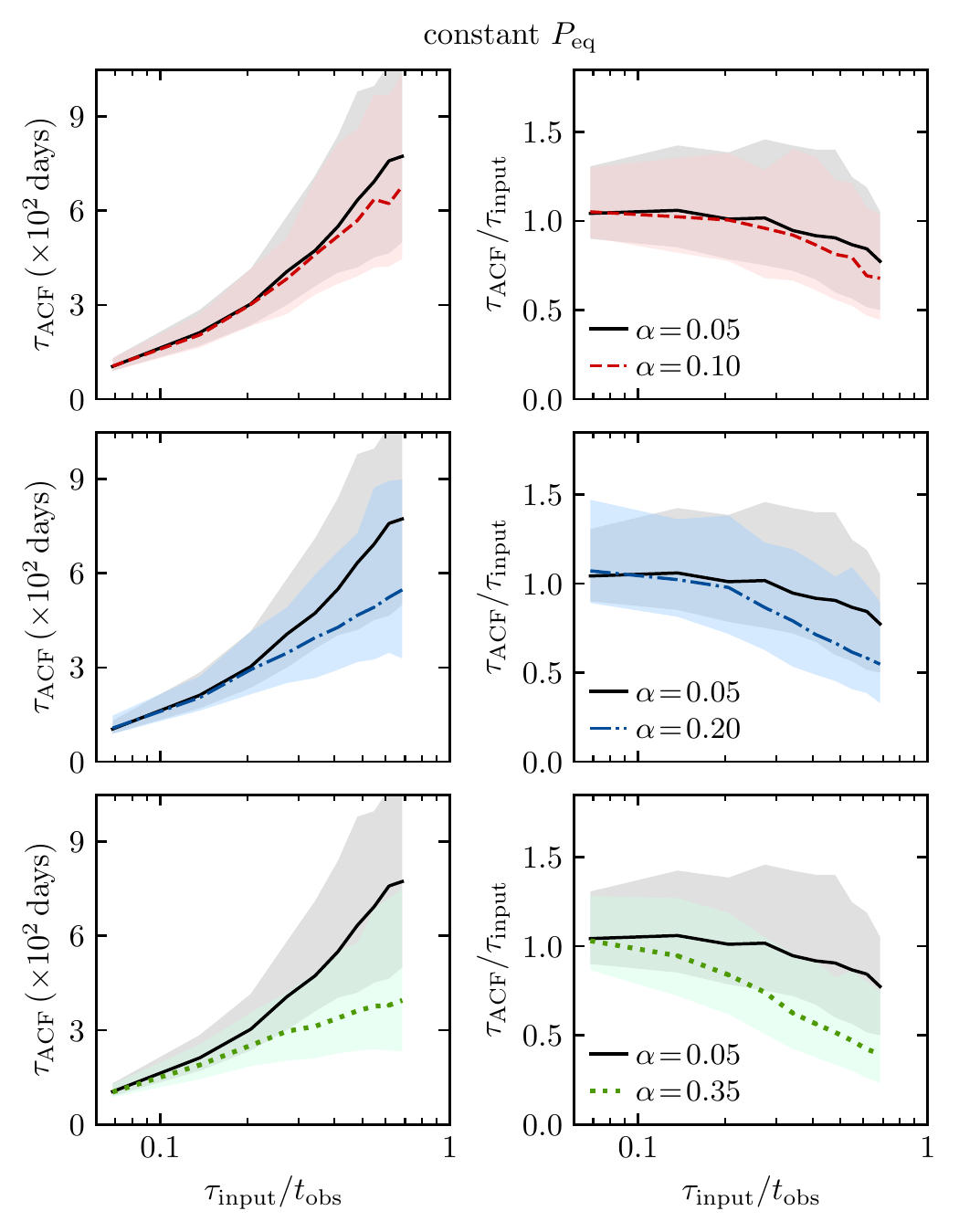}\vspace{-0.6cm}
    \caption{Same as in middle panel of Figure~\ref{fig:rot}, but where the shaded region indicates the 16\textsuperscript{th} and 84\textsuperscript{th} percentiles of the \tacf\ distribution for each \tinput. The solid line indicates the median \tacf\ (\avtacf).}
    \label{fig:rot2A}\vspace{-0.2cm}
\end{figure}

\begin{figure}
    \centering
    \includegraphics[width=\hsize]{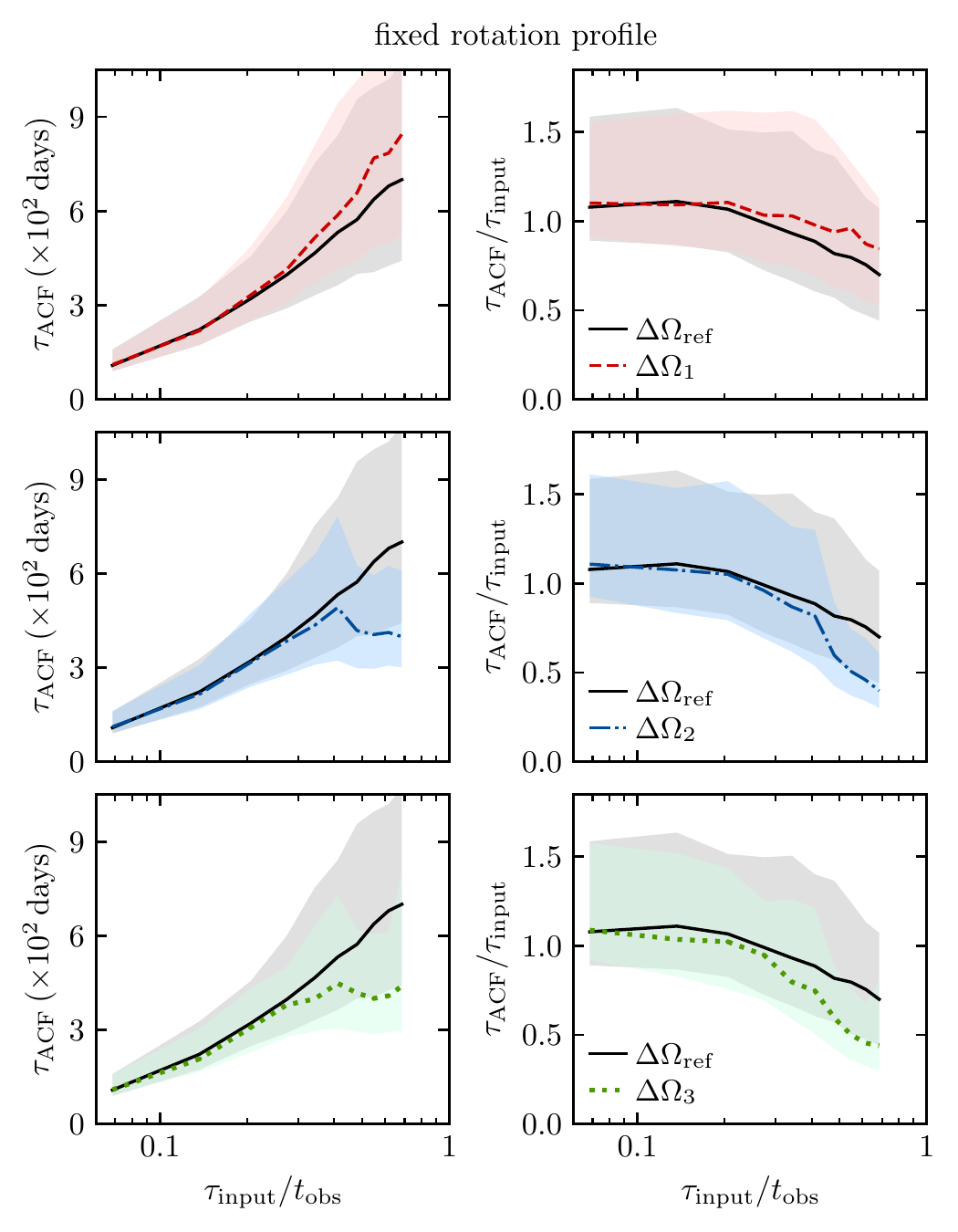}\vspace{-0.6cm}
    \caption{Same as in top panel of Figure~\ref{fig:rot}, but where the shaded region indicates the 16\textsuperscript{th} and 84\textsuperscript{th} percentiles of the \tacf\ distribution for each \tinput. The solid line indicates the median \tacf\ (\avtacf).}
    \label{fig:rot3A}\vspace{-0.5cm}
\end{figure}

\begin{figure}
    \centering
    \includegraphics[width=\hsize]{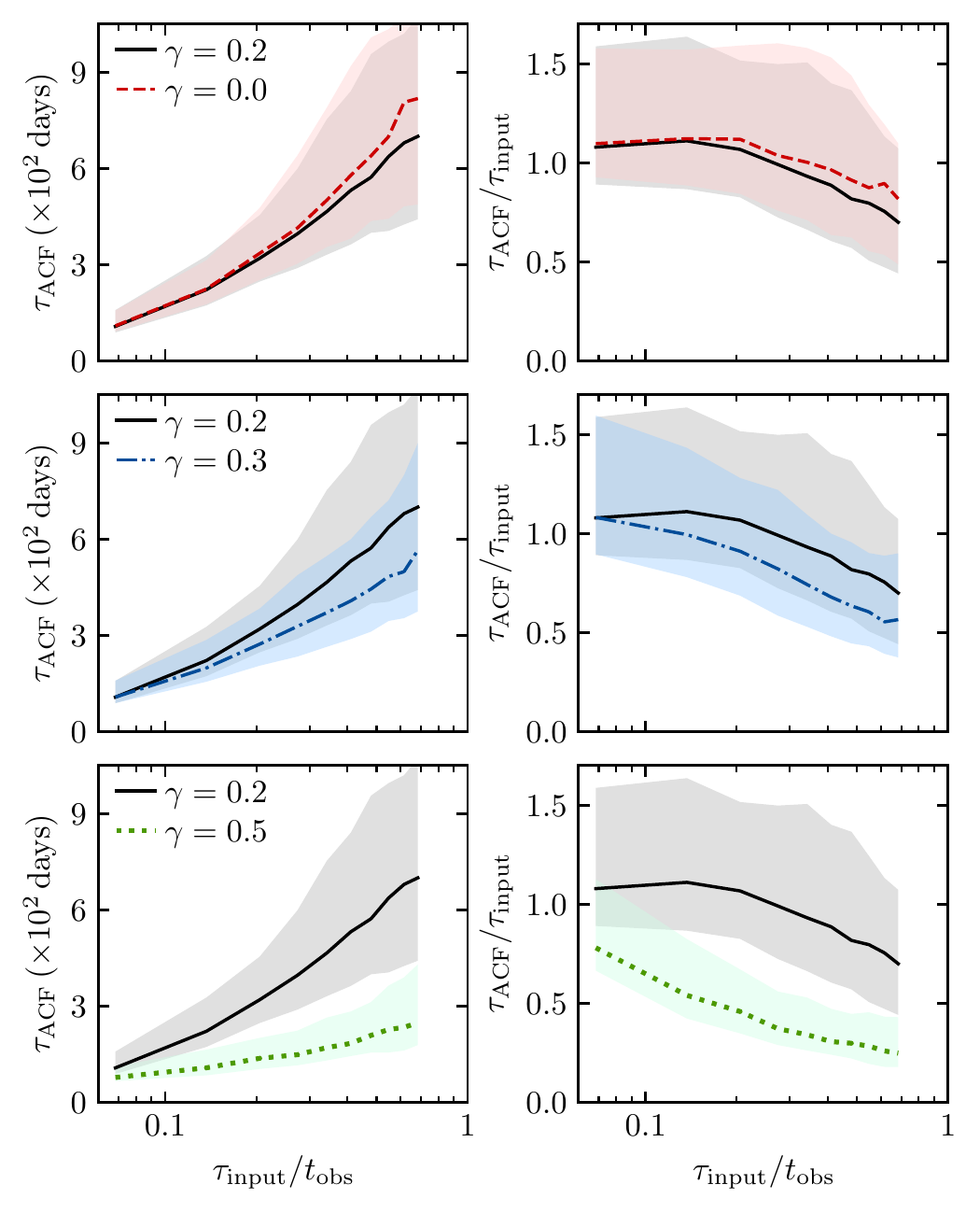}
    \caption{Same as in Figure~\ref{fig:evol}, but where the shaded region indicates the 16\textsuperscript{th} and 84\textsuperscript{th} percentiles of the \tacf\ distribution for each \tinput. The solid line indicates the median \tacf\ (\avtacf).}
    \label{fig:evolA}
\end{figure}

In this section, we present more detailed versions of Figures 4 and 6-10. For easy comparison those figures only show the median values for \tacf. Each of those curves is based on 5000 artificial light curves and there is a significant scatter around the median (Figure~\ref{fig:stdsimul}). As in most of the figures, we compare the results for four different data sets with different properties, we opt to plot only the median \tacf\ in the main figures. Nevertheless, here, Figures~\ref{fig:lengthA}-\ref{fig:evolA} we compare each set with the reference light curves separately and present the 16\textsuperscript{th} and 84\textsuperscript{th} percentiles of the \tacf\ distribution, similarly to Figure~\ref{fig:stdsimul}.

\bsp	% typesetting comment
\label{lastpage}
\end{document}